\documentclass[fleqn,usenatbib]{mnras}
\usepackage{amssymb}

\usepackage{newtxtext,newtxmath}

\usepackage{graphicx}
\usepackage[caption=false]{subfig}
\usepackage{multirow}
\usepackage{rotating}

\usepackage[T1]{fontenc}

\DeclareRobustCommand{\VAN}[3]{#2}
\let\VANthebibliography\thebibliography
\def\thebibliography{\DeclareRobustCommand{\VAN}[3]{##3}\VANthebibliography}

\usepackage{graphicx}	
\usepackage{amsmath}	

\defcitealias{Guzman2016}{Paper~I}
\defcitealias{Guzman2017}{Paper~II}
\defcitealias{Guzman2019}{Paper~III}
\defcitealias{Badnell2021}{Paper~IV}
\defcitealias{PengellySeaton1964}{PS64}
\newcommand{\n}{\mathbf{n}}

\title[H-, He-like recombination spectra -- V]{H-, He-like recombination spectra -- V: On the dependence of the simulated line intensities on the number of electronic levels of the atoms.}

\author[F. Guzm\'an et al.]{
F. Guzm\'an$^{1}$\thanks{E-mail: francisco.guzmanfulgencio@ung.edu (FG)},
M. Chatzikos$^{2}$,
 and G.J. Ferland$^{2}$
\\
$^{1}$Department of Physics \& Astronomy, University of North Georgia, Dahlonega, GA 30597, USA\\
$^{2}$Department of Physics \& Astronomy, University of Kentucky, Lexington, KY 40506, USA\\
}

\date{Accepted XXX. Received YYY; in original form ZZZ}

\pubyear{2022}

\begin{document}
\label{firstpage}
\pagerange{\pageref{firstpage}--\pageref{lastpage}}
\maketitle

\begin{abstract}
This paper presents a study of the 
dependence of the simulated intensities 
of recombination lines from hydrogen and 
helium atoms on the number of $n\ell$-
resolved principal quantum numbers 
included in the calculations. We 
simulate hydrogen and helium emitting 
astrophysical plasmas using the code Cloudy and show that, if not enough $n\ell$-resolved levels are included, recombination line intensities can be predicted with significant errors than can be more than 30\% for H~I IR lines and 10\% for He~I optical lines ($\sim$20\% for He~I IR recombination lines) at densities $\sim1\text{cm}^{-3}$, comparable to interstellar medium. This can have consequences in several spectroscopic studies where high accuracy is required, such as primordial helium abundance determination. Our results indicate that the minimum number of resolved levels included in the simulated hydrogen and helium ions of our spectral emission models should be adjusted to the specific lines to be predicted, as well as to the temperature and density conditions of the simulated plasma. 
\end{abstract}




\section{Introduction}
 
The expansion rate of the early universe might 
have increased with the presence of additional 
neutrino flavors, affecting the neutron-to-
photon ratio, which in turn results in higher 
production of primordial He 
\citep{Steigman2012,Olive2000}. Assessing this 
scenario needs the determination of the helium 
primordial abundances, $Y_p$, to high precision 
\citep{Cyburt2002, Izotov1998}. 

$Y_p$ is commonly obtained by fitting the Big 
Bang Nucleosynthesis (BBN) models to the barion 
density from measurements of the Cosmic 
Microwave Background (CMB). 
\citet{Fields2020,Fields2020b} give a maximum 
likelihood determination of the number of 
neutrino families $N_\nu=2.843\pm0.154$, 
resulting from the Planck mission data. 
Alternatively, spectroscopic determinations of 
the helium primordial abundance focus on 
extragalactic low-metallicity H~II regions, 
where abundances are 
obtained from selected helium and hydrogen line 
ratios \citep{Izotov1998,Aver2015,Valerdi2019}. 
$Y_p$ has also been obtained from observations 
of the intergalactic medium \citep{Cooke2018}. 
High statistics on the observed targets help to 
minimize instrumental and observational errors 
\citep{Izotov2013}, although the introduction of 
systematic error in the observations can be 
dominant \citep{Aver2010,Aver2011,Peimbert2016}. 
Other significant uncertainties reside in the 
values of the probabilities of the electron 
collisional processes with the hydrogen and 
helium ions. 
    
Comparisons with synthetic spectra is then 
essential to model helium abundances. Modeling 
has its own sources of uncertainty. For example, 
a complete description of each species' infinite 
number of quantum levels is not available to any 
computer. For moderate densities 
($\n=10^4\text{cm}^3$), high Rydberg levels can 
be considered in local thermodynamic equilibrium 
(LTE), where the collisions balance the electron 
populations statistically. As electron density 
decreases, more and more levels must be 
explicitly considered, eventually elevating the 
calculation times and computational memory to 
prohibitive numbers. 

Several techniques have been used to bypass 
this difficulty. \citet{Seaton64.Hpops} 
describes the equations that lead to high 
$n$ population departures from LTE. 
\citet{Brocklehurst1972} simplified the 
collisional-radiative (CR) matrix using a 
"condensation" scheme, interpolating high-$n$ 
departure coefficients from a number of 
representative levels. For optically thin 
plasmas, \citet{Burgess1976} created CR 
coefficients to describe the fast "ordinary" 
collisional radiative processes in function of 
the slower metastables, which are assumed to be 
in pseudo static equilibrium. 

A related problem is the treatment of $n\ell$ 
sub-levels that should be accounted for 
explicitly in order to 
correctly describe emission lines from 
electronic transitions between $n\ell\to 
n^\prime\ell^\prime$ sub-shells. Every $n$-shell has $n^2$ sub-shells, which increases 
the computing operations in $\sim n^4$ 
\citep[][hereafter Paper III]{Guzman2019}. The standard approach 
consists in setting a minimum $n$ for which 
proton collisions dominate over radiative 
transitions, and the populations of the $n$-
shells will be $\ell$-mixed 
\citep[hereafter PS64]{PengellySeaton1964}. Over this limit, 
density and $\ell$-changing collisional rates 
are assumed to be large enough for the $\ell$-shells 
to populate statistically with electrons 
so $n$-shells can be treated as unresolved in $
\ell$ \citep[also known as {\it collapsed}; see]
[]{C17}. Under this limit, $n$-shells will be 
resolved. The number of resolved levels should 
be small enough to avoid long calculation times 
but high enough to account for cascade electrons 
that will contribute to line emission. 

In this paper, we will use the self-consistent 
spectral code Cloudy \citep{C23} to analyze the 
influence of the number of $n\ell$-resolved 
states on the predicted intensities of the 
hydrogen and helium lines. We aim to signal 
a potential source of uncertainty in the models 
and provide a method to avoid it. We show that 
these corrections are essential in studies 
that require a great accuracy, such as the 
primordial helium determinations mentioned 
above. Moreover, they could also be important in 
calculating helium abundances in solar 
metallicity H~II regions with low to moderate 
electron density \citep[see, for example,][]
{MendezDelgado2020}. 

This paper is the fifth in a series where we 
carefully review the theory and methods for 
predicting the recombination spectra of H- and 
He-like ions. Section \ref{sec:methods} 
reviews the atomic data employed in the code 
Cloudy. Section \ref{sec:critdens} analyzes the 
models' dependence on the number of resolved and 
unresolved levels included. Section 
\ref{sec:hhesims} describes our benchmark 
simulation of hydrogen and helium plasmas. 
Section \ref{sec:results} presents our results, 
followed by a discussion in section 
\ref{sec:discussion}. 

\section{Atomic Data}
\label{sec:methods}

The code Cloudy uses a collisional-radiative 
approach for H-like and He-like ions. All other 
ions are treated using a two-level calculation, 
where ionization-recombination balance is 
obtained by only considering ionization from the 
ground state of the species under study and 
where electrons recombined to all excited states 
will eventually decay to the ground state. For 
these ions, emissions from low-lying states are 
assumed not to be affected by the ionization/
recombination processes due to the largely 
different time-scales for ionization/
recombination and excitation \citep[see][for a 
discussion on the limits of this approximation]
{C17}. In contrast, coupling of the H-like and 
He-like iso-electronic sequences' excited levels 
with the continuum can make an impact in the 
optical and infrared emission lines. 

The atomic data used in this work for H-like and 
He like iso-sequences are summarized below. 

\subsection{Bound-free Radiative Transitions}

We calculate radiative recombination coefficients 
for H-like ions using the Milne relation (detailed 
balance) from the photoionization cross sections 
\citep{Brocklehurst1971}.

\citet{HS98} provide radiative photoionization 
data for  atomic helium for $n\leq25$ and $
\ell\leq3$ or $n>25$ and $\ell\leq 1$. For higher $
\ell$'s, these authors scale the recombination 
coefficients to hydrogen. 

Data for photoionization from the ground level of 
He-like ions are obtained from the fits of 
\citet{Verner1996b}, while excited levels up to 
$n=10$ from lithium through calcium are obtained 
from the TopBase database at the Opacity 
project\footnote{\url{https://cdsweb.u-strasbg.fr/topbase/topbase.html}} \citep{Cunto1993}. 
Photoionization coefficients from higher $n$'s are 
roughly approximated to be hydrogenic.

\subsection{Bound-Bound radiative transitions}

We calculate the Einstein radiative de-excitation 
coefficients as a function of the oscillator 
strengths 
$f_{nl,n^\prime l^\prime}$ \citep[see][eq. 10.17]{Martin2006}:

\begin{equation}
A_{n\ell,n^\prime \ell^\prime} = \frac{2\pi e^2}{m_ec^3\epsilon_0}\nu^3
\frac{g_{n^\prime \ell^\prime}}{g_{n\ell}}f_{n\ell,n^\prime \ell^\prime}\,,
\label{eq:Aval}
\end{equation}

\noindent where $m_e$, $e$, and $c$ are the 
electron mass, the electron charge, 
and the speed of light respectively, $\epsilon_0$ 
is the permittivity of free 
space, $\nu$ is the frequency of the transition, 
and $g_{nl}$ is the 
statistical weight of the $nl$ level. The 
oscillator strengths are calculated as a 
function of the radial integral as 
\citep{Mansky2006}:

\begin{equation}
  f_{n\ell,n^\prime \ell^\prime} = \frac{\hbar\omega}{3R_\infty}
  \frac{\max(\ell,\ell^\prime)}{2\ell+1}\left|R_{n\ell}^{n^\prime \ell^\prime}\right|\,,
    \label{eq:oscstrength}
\end{equation}

\noindent with $\omega=2\pi\nu$ the angular 
frequency of the transition in
s$^{-1}$, and $R_\infty$ the Rydberg constant for 
infinite nuclear mass. The
radial integrals $R_{nl}^{n^\prime l^\prime}$ are 
calculated using the
recursion relation of the hyper-geometric 
functions \citep{Hoangbinh1990}, which
are part of the exact solution provided by 
\citet{Gordon1929}.

In Cloudy's collisional-radiative approach, high 
Rydberg levels are unresolved
on the angular momentum. Those levels are dubbed 
\emph{collapsed} and
are assumed to have statistical populations in the 
$\ell$-subshells \citep{C17}.
For them, we use the formula of 
\citet{Johnson1972}:

\begin{equation}
  A_{n,n^\prime} = \frac{32}{3\sqrt{3}\pi}n\sum_{i=0}^2 \frac{g_i(n)}{i+3},
  \label{eq:J72A}
\end{equation}

\noindent where $g_i(n)$ are the factors of a 
polynomial approximation
($g(n,x)=g_0(n) +g_1(n)x^{-1}+g_2(n)x^{-2}$) of 
the bound-free Gaunt factor
\citep{MenzelPekeris35}.

\subsection{$\ell$-changing collisions}
\label{sec:lchanging}

$\ell$-changing or $\ell$-mixing collisional 
transitions change the angular momentum $\ell$ of 
an electron of a given $n$-shell by proton Stark 
field mixing \citep{PengellySeaton1964,Vrinceanu2001}. The transition 
probabilities increase at lower temperatures due 
to a longer interaction of the proton electric 
field with the target atom shells \citep[]
[hereafter Paper I]{Guzman2016}. $\ell$-changing 
collisions produced by
slow charged heavy particles, such as protons or 
alpha particles, are dominant.

 We use an improved version of the approach of 
 \citetalias{PengellySeaton1964}, 
 which we named PS-M, developed by 
 \citet[hereafter Paper II]{Guzman2017}, and 
 extended to non-degeneracy cases (such as 
 hydrogen-like helium ions) and low-temperature/
 high-densities cases, where the classical PS64 
 produced nonphysical results, in 
 \citet[hereafter paper IV]{Badnell2021}. We 
 use the formulas recommended in the equation 
 (9) and (12) of \citetalias{Badnell2021}. Comparison of these 
 results with the ones from PS64, used by HS87, 
 are given in table 1 of \citetalias{Badnell2021} for $n=30$, 
 showing a ratio between the 
 two theories of $\sim30\%$ for $\ell=4 \to 
 \ell^\prime =3$ and $\sim 4\%$ for $\ell=29 \to 
 \ell^\prime =28$. 

\subsection{n-changing collisions}
\label{sec:nchanging}

For hydrogen atoms, energy changing electron-impact de-excitation 
effective rate coefficients up to $n=5$ are
taken from the R-Matrix with
pseudo-states results in table 2 of 
\citet{Anderson2000, Anderson2002}. For other 
H-like ions, fits from \citet{Callaway1994} and 
\citet{Zygelman1987} are used for $n=1 - 2$ collisions. For 
helium atoms, effective rate coefficients for $n\leq 5$ are 
taking from the convergent close coupling (CCC) calculations from 
\citet{Bray2000}. For He-like ions except Fe$^{24+}$, $n=1,2 \to 
n=2$ electron impact collisional transitions are taken from the 
fits and tabulations of \citet{Zhang1987} based on their own 
calculations. For He-like iron, we prefer the more accurate 
results of \citet{Si2017} using the independent process and 
isolated resonances approximation using distorted waves (IPIRDW).

For higher principal quantum numbers, straight 
trajectory Born approximation is
used \citep[equation 8.30 in][]{Lebedev1998}. 
These rates are
angular momentum unresolved $n\to n^\prime$. To 
obtain $n\ell\to n^\prime \ell^\prime$ resolved
effective rates, we average over the statistical 
weight of the initial and final levels:

\begin{equation}
  q(n\ell\to n^\prime \ell^\prime)=\frac{\left(2\ell^\prime+1\right)}{{n^\prime}^2}q(n\to n^\prime)\,.
  \label{eq:Wrates}
\end{equation}

\subsection{Collisional Ionization and Three-Body Recombination}
\label{sec:ion}

We use \citet{Vriens1980} 
formulae for collisional ionization. Three-body
recombination values are obtained from detailed 
balance. 

\section{Dependence of the calculations on the number of atomic levels}
\label{sec:critdens} 

The more energy levels included in the 
calculations, the more accurate the description of 
recombination electrons cascading down to the 
lower levels to produce recombination lines. 
Additionally, the separate treatment of $\ell$-
subshells when they are not statistically 
populated, e.g., $\ell$-changing collisions do 
not dominate over the radiative probabilities, 
leads to a precise calculation of the line 
intensities. In modeling, choosing the correct 
number of resolved levels can be tricky. In this 
section, we briefly show how to calculate the 
limiting $n$ values for which the non-inclusion of 
$\ell$-resolved levels in the spectral simulations 
can make an impact.

\subsection{Continuum Coupling}

\begin{figure}
    \centering
    \includegraphics[width=0.5\textwidth]{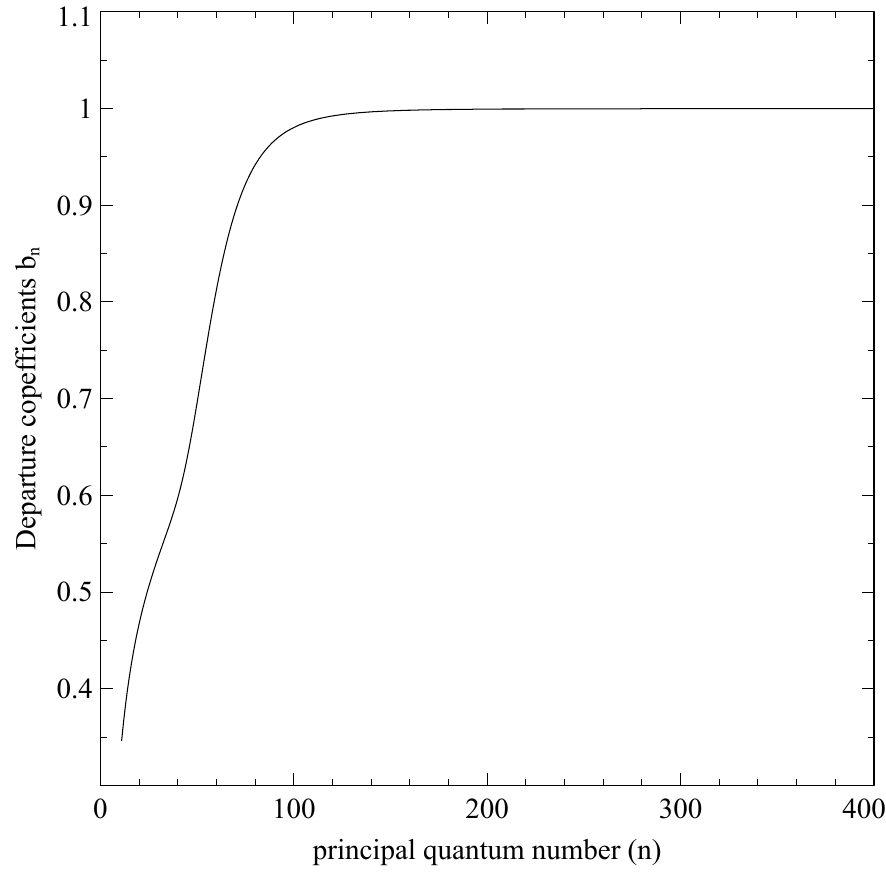}
    \caption{Departure coefficients for collapsed 
    levels of a hydrogen gas at $\n_\text{H} = 10^4 
    \text{cm}^{-3}$ ionized by a monocromatic 
    radiation of 2Ryd.}
    \label{f:depcoef}
\end{figure}

To account for the offset that the computation of 
a finite number of levels of a model ion can
have in the final calculation we ``top off'' the 
highest $n$-shell
\citep{Bauman2005} by increasing its ionization 
rate by one hundred times, forcing it to LTE with 
the continuum. The immediate lower levels will 
then receive the electron cascade from the 
continuum, coupling through radiative decay and 
mostly from $n$-changing collisions. However, this 
treatment is inaccurate for models with $n<100$ at 
intermediate densities
\citepalias[see appendix in][]{Guzman2019}.
At higher densities, continuum-lowering effects 
\citep{Bautista00} can reduce
the top-off errors. As a result, high $n$-shells 
tend to approach LTE with the continuum and 
between themselves due to the high excitation 
rates \citepalias{Guzman2019}. We define departure 
coefficients $b_n$ as the departure of the shell 
population from LTE:

\begin{equation}
\n_n = \n_n^{\text{LTE}} b_n = \n_e\n_+\left(\frac{h^2}{2\pi m_e kT_e}\right)^{3/2}\frac{g_n}{2}e^{\frac{E_n}{kT_e}} b_n,    
\end{equation}

\noindent where $\n_+$, $\n_e$, $m_e$, and $T_e$ are the 
parent ion density, electron density, electron mass, and electron 
temperature, respectively; $h$ is the Planck 
constant, $k$ is the Boltzmann constant, $E_n$ is 
the $n$-shell binding energy, and $g_n$ is the 
statistical weight of the shell $n$. When the 
departure coefficients are close to 1, the levels 
are at LTE with each other and with the continuum. 
In Figure \ref{f:depcoef}, we show departure 
coefficients for a slab of gas at density 
$\n_\text{H} = 10^4 \text{cm}^{-3}$ under a 
monochromatic ionization source at 2 Ryd. In the 
Figure, the departure coefficients for high $n$ 
smoothly approach unity. A calculation must 
include enough $n$-shells at LTE to 
describe the electron cascade from higher 
levels accurately.
This method works well at intermediate hydrogen 
densities of $\n_{\text{H}} \lesssim 10^5 \text{cm}
^{-3} $ when the principal quantum number of the 
last $n$-shell is $n_{\text{last}}\gtrsim100$ 
\citepalias{Guzman2019}. At higher densities, the 
effects of continuum lowering can reduce errors 
\citep{Bautista00}. 

More rigorous approaches have been used in the 
literature. \citet{Hummer1987} CR matrix 
condensation 
\citep{Burgess1969,Burgess1976,Brocklehurst1970} 
interpolates and extrapolates high-$n$ departure 
coefficients to a smooth function. 

\subsection{ $\ell$-changing critical densities}

 The line emissivity for a transition from the 
 upper level $n\ell$ to the lower level 
 $n^\prime\ell^\prime$ is given by:

 \begin{equation}
     \epsilon_{n\ell\to n^\prime\ell^\prime} = 
     \n_{n\ell} A_{n\ell,n^\prime\ell^\prime} 
     \frac{h\nu}{4\pi},
     \label{eq:emiss}
 \end{equation}

 \noindent where $\nu$ is the frequency of the 
 transition, $\n_{n\ell}$ is the density of the 
 upper level, and 
 $A_{n\ell,n^\prime\ell^\prime}$ is the Einstein 
 coefficient value of that transition. It is 
 important to obtain precise equilibrium 
 populations on the upper levels of the 
 transitions to obtain accurate line 
 emissivities. If these levels are not $\ell$-
 mixed, they must be resolved in their 
 corresponding $\ell$-shells. $\ell$-changing 
 collisions are more effective the higher the 
 principal quantum numbers $n$ are, the effective 
 coefficients varying as 
 $q_{\ell\ell^\prime} \sim n^4$ if $\ell\ll n$ 
 \citepalias{PengellySeaton1964}. At high $n$, $\ell$-
 changing is dominant over $n$-changing 
 transitions, and the $n$-shells are 
 statistically populated in all the $n\ell$-
 subshells,

\begin{equation}
 \n_{n\ell} = \frac{2\ell + 1}{n^2} \n_n,
\label{eq:nstat}
\end{equation}

\noindent where $n_{n\ell}$ is the density of 
the $n\ell$-subshell. The competing process is 
spontaneous decay. The radiative lifetime is $
\tau_{n\ell} \sim n^5$, and so it can be 
compared with the collisional redistribution 
lifetime, 

\begin{equation}
    \tau_\text{coll} = \left(q_{\ell\ell^\prime}
    \n_\text{coll}\right)^{-1},
    \label{eq:colltime}
\end{equation}

\noindent where $n_\text{coll}$ is the collider 
density. We define the critical density at the 
point where $\tau_\text{coll}=\tau_{n\ell}$ as

\begin{equation}
    \n^\text{crit}_\text{coll} = 
    \left(\tau_{n\ell}
    q_{\ell\ell^\prime}\right)^{-1}.
    \label{eq:critdens}
\end{equation}

In astrophysical plasmas, such as H~II regions, 
we can assume $\n_\text{coll} \sim \n_\text{H} 
\sim \n_e$ because hydrogen atoms are about 90\% 
of all particles, although Helium ions and alpha 
particles could contribute to $\ell$-changing 
collisions, decreasing critical densities in eq. 
\eqref{eq:critdens}, as He abundance is usually 
around 10\%. In the rest of this paper, we will 
assume that $\n_\text{coll} \sim \n_\text{H}$ 
unless otherwise specified. 

\begin{figure}
    \centering
    \includegraphics[width=0.5\textwidth]{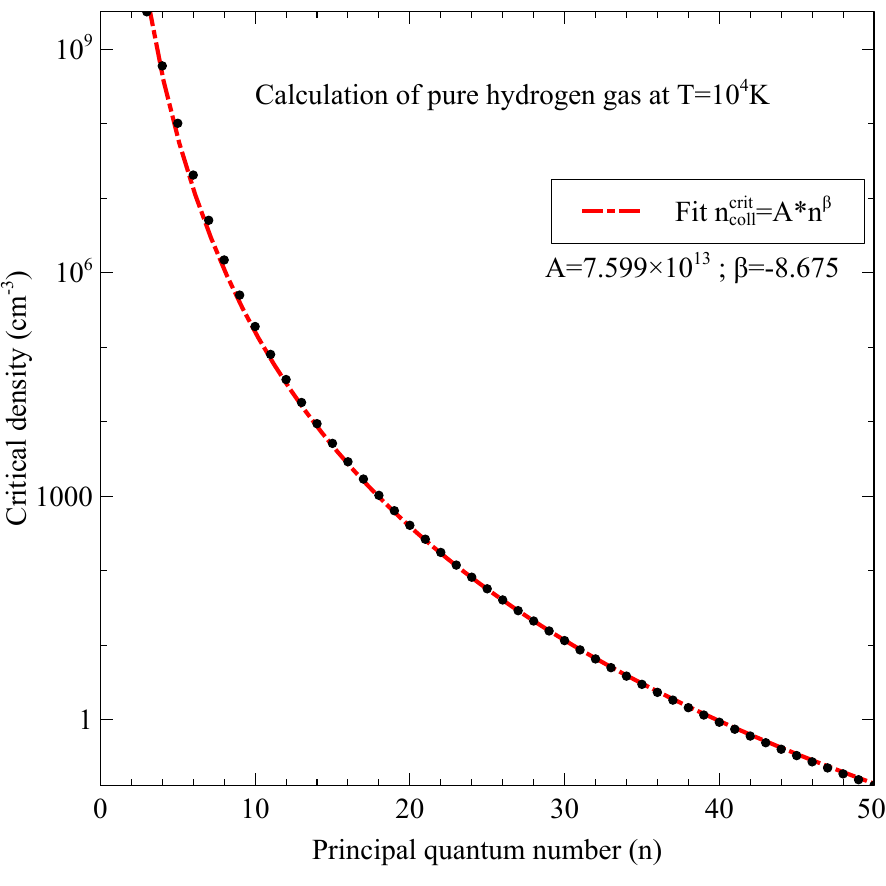}
    \caption{$\ell$-mixing critical densities as 
    a function of the principal quantum number. 
    The dots correspond to our results using 
    equation (\ref{eq:critdens}). Fits give a 
    power law $\sim n^{-9}$. }
    \label{f:ncritfitT4}
\end{figure}

The total radiative lifetime depends on the 
principal quantum number $n$ as $\tau_{n\ell} = 
A_{n\ell}^{-1} \propto n^{5}$, and, together 
with the dependence of the collisional rates on 
$n$, will make $\n^\text{crit}_\text{coll} 
\propto n^{-9}$. In Figure \ref{f:ncritfitT4}, 
we fitted the critical densities obtained for 
pure hydrogen gas at $T = 10000$K. Our fit, 
$\n^\text{crit}_\text{coll}=An^\beta$ with 
$A=7.6\times10^{11}\text{cm}^{-3}$ and $\beta = 
8.68$, coincides with our estimations.

\begin{figure}
    \centering
    \includegraphics[width=0.5\textwidth]{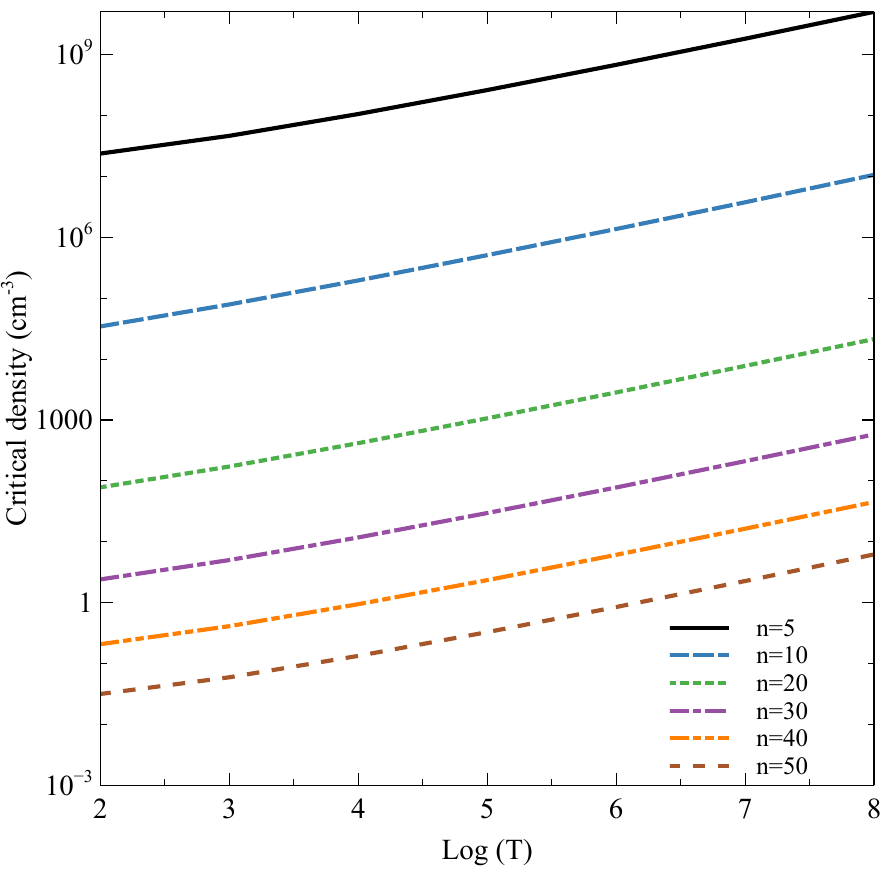}
    \caption{$\ell$-mixing critical densities as 
    a function of the temperature for different 
    principal quantum numbers of hydrogen atoms. 
    Note that the $\ell$-changing rates 
    $q_{\ell\ell^\prime}$ decrease with the 
    temperature because at higher kinetic 
    energies the time in which the projectile 
    contributes to the Stark mixing is reduced 
    \citepalias[see fig. 1 in][]{Guzman2016}. 
    This will produce an increase of the 
    critical densities at higher temperatures. }
    \label{f:ncritvsT}
\end{figure}

This simple fit produces a method to estimate 
the minimum principal quantum number to be $
\ell$-resolved. However, it is necessary to 
account for the temperature dependence of the 
collisional rates $q_{n\ell}$, which can 
significantly vary the critical densities. In 
Figure \ref{f:ncritvsT}, the critical densities 
range over two orders of magnitude between 
$T=100$K to $T=10^8$K. For example, at electron 
density $\n_e=10^4\text{cm}^{-3}$, the level 
$n=30$ can be $\ell$-mixed at $T_e=100$K, but it 
will not be so at $T_e>10^5$K. We have fitted 
the dependence of the critical density with the 
temperature to $\n^\text{crit}_\text{coll} = 
B+CT^\gamma$, where $\gamma\approx 0.44$, and 
$B$ and $C$ depend on the specific principal 
quantum number. 

\begin{figure}
    \centering
    \includegraphics[width=0.5\textwidth]{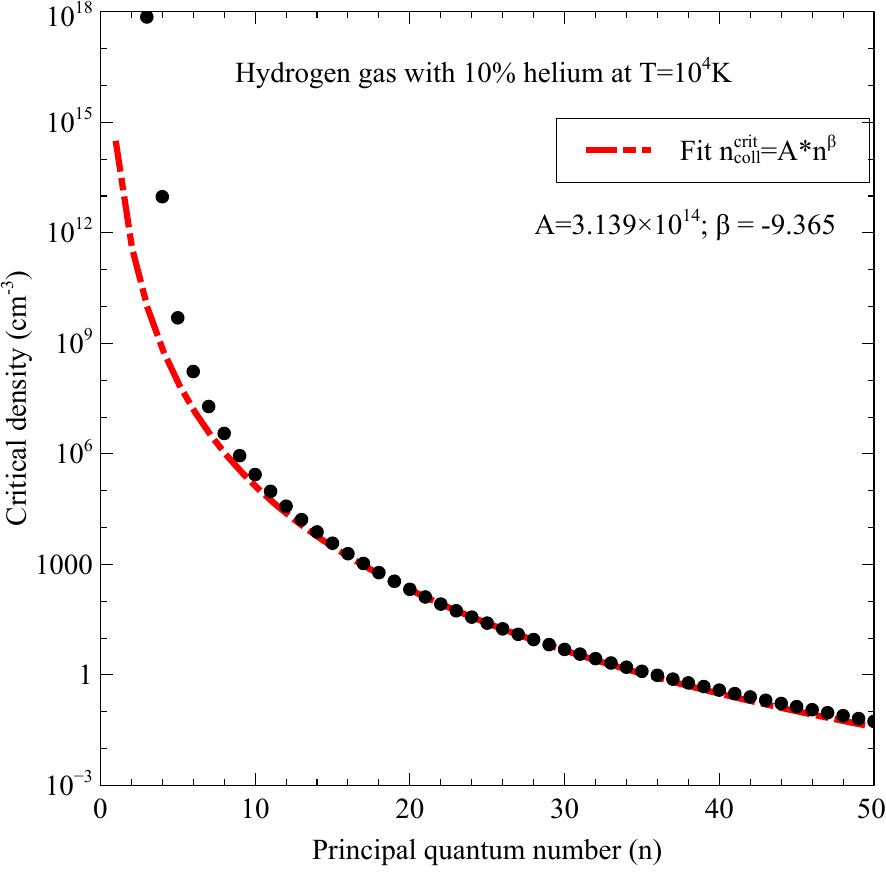}
    \caption{ He~I $\ell$-mixing critical 
    densities as a function of the principal 
    quantum number for a hydrogen and helium 
    mixed plasma (see text for details). The 
    dots correspond to our results using 
    equation (\ref{eq:critdens}). Fits give a 
    power law $\sim n^{-9}$ that only works for 
    high principal quantum numbers.}
    \label{f:ncrithe}
\end{figure}

\begin{figure*}
    \centering
        \includegraphics[width=0.4\textwidth]{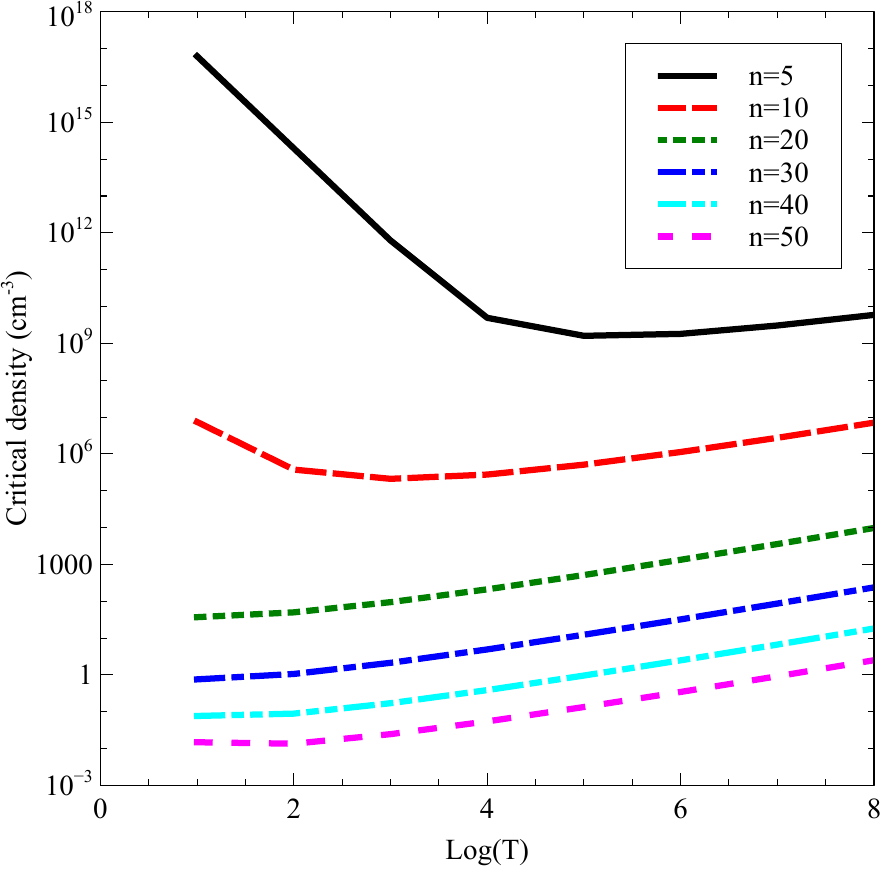}
\qquad
        \includegraphics[width=0.4\textwidth]{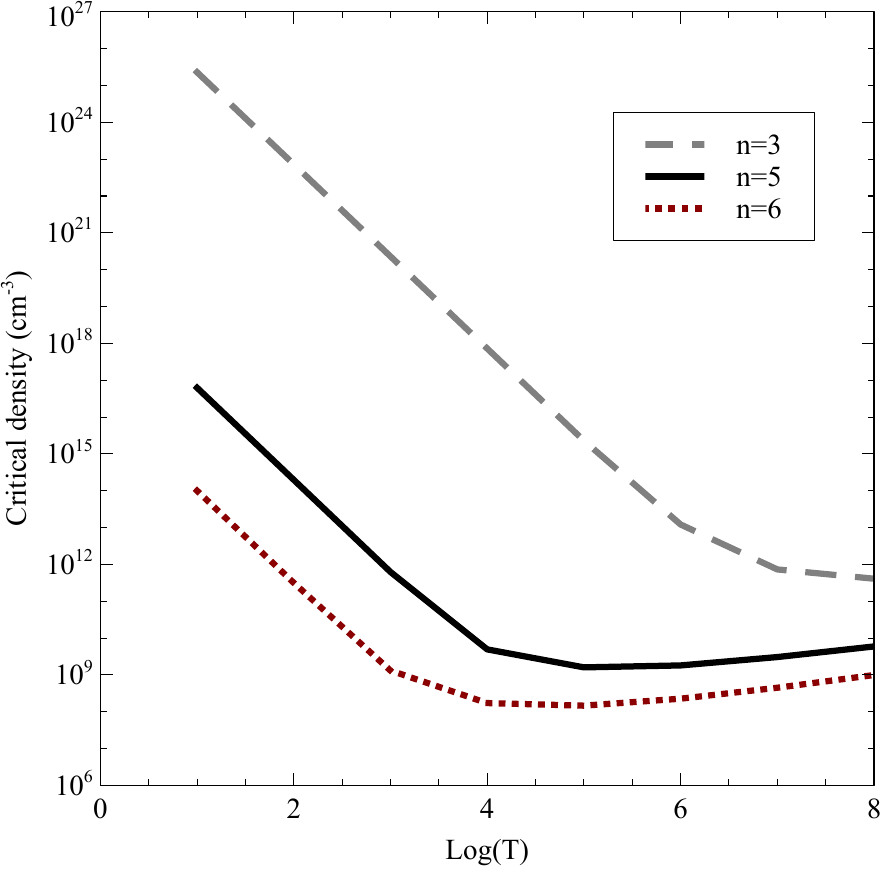}
    \caption{$\ell$-mixing critical densities as 
    a function of the temperature for different 
    principal quantum numbers of helium atoms. 
    Left: high $n$-shell. Right: $n=3$, $n=5$, 
    and $n=6$. The extra cut-off term due to 
    the non-degeneracy of helium levels make 
    the $\ell$-changing rates 
    $q_{\ell\ell^\prime}$ have a different 
    shape than in the hydrogen case 
    \citepalias{Guzman2017}. This non-
    degeneracy cut-off is stronger for low $n$ 
    and low temperatures. }
    \label{f:ncrithevsT}
\end{figure*}

This situation is slightly more complicated for 
helium atoms, as $\ell$-changing collision 
probabilities depend on an extra cut-off of the 
probability at large impact parameters due the 
broken degeneracy of the $\ell$-shells 
\citepalias{Guzman2017}. The cut-off acts at low 
temperatures, increasing the critical density, 
lowering the rate coefficients, and making them 
have a more complicated temperature dependence 
than in the hydrogen case. This effect is mainly 
happening at low $n$'s, with a larger 
degeneracy. In Figures \ref{f:ncrithe} and 
\ref{f:ncrithevsT}, critical densities are 
plotted as a function of the principal quantum 
number and temperature. The dependence of the $
\ell$-changing rate coefficients on the electron 
temperature reflects on the critical densities. 
In Figure \ref{f:ncrithe}, the critical densities 
of high Rydberg helium levels still depend on 
the principal quantum number as $\n^\text{crit}
_\text{coll} \sim n^{-9}$.

$n$-shells are at critical densities when 
outbound radiative decay and inbound $\ell$-
changing collisions are equally competing. 
Thus, these levels should not be expected to be 
statistically populated in the $\ell$-subshells 
until collisions strongly dominate, making it 
necessary to resolve $n$-shells until the 
critical densities are well below the density 
of the simulated plasma. A rule of thumb
is to keep resolved levels up to 10+$n$, with 
$n$ the principal quantum number of the level 
for which the critical density equals the 
density of the plasma. However, as seen in Figure 
\ref{f:ncritfitT4}, critical densities are more spaced for lower levels. 
Therefore, this rule will work better 
at higher densities, where a relatively lower 
number of n-shells needs to be resolved. 
Specifically, for densities below $
\sim1\text{cm}^{-3}$, levels well over 10+$n$ 
must be kept $\ell$-resolved.

\section{Hydrogen and helium ionized plasma simulations}
\label{sec:hhesims}

To quantify the error in the 
line intensities produced by an insufficient 
atom description, we carried simulations using 
the last development version of the spectral code Cloudy 
\citep[master 13108b32,][]{C23}, varying the number of resolved 
levels. We restrict ourselves to a slab of pure 
hydrogen gas or a mixture of hydrogen and 
helium to understand the effect of the size of 
atoms. We have illuminated our plasma with a 
monochromatic radiation of 2 Ryd, enough to 
ionize helium and hydrogen atoms. We set the 
ionization parameter as

\begin{equation}
    U = \frac{\phi_\nu}{\n_\text{H}c} = 0.01,
\end{equation}

where $\phi_\nu$ is the photon 
flux\footnote{This is usually the flux of 
hydrogen ionizing photons but, since we are 
using a monochromatic source, all photons are 
ionizing hydrogen.} at the surface of the cloud 
in $\text{ergs}\cdot\text{s}^{-1}\cdot\text{cm}
^{-2}$, $\n_\text{H}$ is the hydrogen density, and $c$ 
is the speed of light. This value will ensure 
enough photons to ionize our gas at all 
densities\footnote{In this case, $\n_e=\n_\text{H}$ for a pure hydrogen gas and $\n_e=1.1\n_\text{H}$ for a 10\% mix of He (section \ref{sec:herec})}. We restrict the thickness of our 
slab to 1cm to minimize optical depth and 
reduce the computational time. We assume case B 
\citep{Baker1938} approximation, where Lyman 
emission lines (except Ly$\alpha$) are 
reconverted into higher series. Strict Case B 
approximation is accurate at densities 
$\n_e\leq10^8\text{cm}^{-3}$ \citep{Hummer1987}. 
We relax strict Case B approximation by 
explicitly accounting for collisions from $n=2$ 
levels, which allows to carry this 
approximation up to higher densities.
 
\begin{table*}[p]
    \centering
        \caption{ H I recombination lines corresponding to the decays of the first eight series (starting with Balmer, $n=2$) up to $n=10$ that have been tested in our pure hydrogen plasma model at $T=10^4$K. The columns in the right contain the maximum resolved principal quantum number $n$ at which the line intensity converge for less than 5\% and 1\% with respect to the models with maximum resolved levels $n^\text{res} = n-5$ (convergence ratios start to be calculated at $n=15$). We show these at two relevant hydrogen densities of $\n_\text{H} = 1\text{cm}^{-3}$ (corresponding to the interstellar medium) and $\n_\text{H} = 10^4\text{cm}^{-3}$ (H~II regions). For each density, the maximum percentage difference induced by changing the maximum resolved level from  $n = 70$ to $n = 10$ is also shown (columns "Diff.").}
    \label{tab:linesh}
    \begin{tabular}{c|c|c|c|c|c|c|c|c}
       $\lambda$ & transition & comments & \multicolumn{3}{|c|}{Convergence at $\n_\text{H}=1\text{cm}^{-3}$} & \multicolumn{3}{|c|}{Convergence at $\n_\text{H}=10^4\text{cm}^{-3}$}\\
      &&& n ($<$5\%) & n ($<$1\%) & Diff. & n ($<$5\%) & n ($<$1\%)& Diff. \\
      \hline
      \hline
  1215.67\AA & $1\,^2S -   2\,^2P$ & Ly$\alpha$ & 15 & 15& 0.9\%& 15 & 15&0.21\%\\
   3797.90\AA & $2\,^2S - n=  10$ & &15 & 15& 1.81\%& 15& 15& 1.43\%\\ 
   3835.38\AA & $2\,^2S - n=  9$ & &15 & 15& 1.84\% & 15&15 & 1.45\%\\
   3889.05\AA & $2\,^2S - n=  8$ & &15 & 15& 1.82\%& 15& 15& 1.43\%\\
   3970.07\AA & $2\,^2S - n=  7$ & &15 & 15& 1.78\%& 15& 15 & 1.39\%\\
   4101.73\AA & $2\,^2S - n=  6$ & &15 & 15& 1.64\%& 15& 15 & 1.28\%\\ 
   4340.46\AA & $2\,^2S - n=  5$ & &15 & 15& 1.32\%& 15& 15& 1.03\%\\
   4861.32\AA & $2\,^2S - n=  4$ & H$\beta$ & 15 & 15& 0.46\%& 15& 15& 0.47\%\\
   6562.80\AA & $2\,^2S - n=  3$ & H$\alpha$ & 15 & 20 & 2.75\%& 15 & 20& 2.02\%\\
   9014.91\AA & $n=  3 - n=  10$ & & 15 & 15& 0.10\%& 15& 15& 0.10\% \\ 
   9229.02\AA & $n=  3 - n=  9$ & & 15 & 15& 0.16\%& 15& 15& 0.10\% \\ 
   9545.97\AA & $n=  3 - n=  8$ & & 15 & 15& 0.51\%& 15& 15& 0.37\%\\ 
   1.00494$\mu$m & $n=  3 - n=  7$ & & 15 & 15& 1.05\%& 15& 15& 0.77\%\\ 
   1.09381$\mu$m & $n=  3 - n=  6$ && 15 & 20 & 1.99\%& 15& 20& 1.48\%\\ 
   1.28181$\mu$m & $n=  3 - n=  5$ && 15 & 20& 4.06\%& 15& 20& 3.00\% \\
   1.73621$\mu$m & $n=  4 - n=  10$ & & 15 & 20& 3.23\%& 15 & 20& 2.41\%\\
   1.81741$\mu$m & $n=  4 - n=  9$ & & 15 & 20& 3.87\%& 15& 20& 2.89\%\\  
   1.87510$\mu$m & $n=  3 - n=  4$ & & 15 & 30& 10.30\%& 15& 25& 7.61\% \\
   1.94456$\mu$m & $n=  4 - n=  8$ & & 15 & 25& 4.79\%& 15& 20& 3.57\%\\  
   2.16553$\mu$m & $n=  4 - n=  7$ & & 15 & 25& 6.33\%& 15& 25& 4.70\%\\ 
   2.62515$\mu$m & $n=  4 - n=  6$ & & 15 & 30& 9.40\%& 15& 25& 6.92\%\\
   3.03837$\mu$m & $n=  5 - n=  10$ & & 15 & 25& 7.71\%& 15 & 25& 5.72\%\\
   3.29609$\mu$m & $n=  5 - n=  9$ & & 15 & 25& 9.03\%& 15 & 25& 6.69\%\\
   3.73954$\mu$m & $n=  5 - n=  8$ & & 20 & 30& 11.10\%& 20 & 25& 8.69\%\\
   4.05115$\mu$m & $n=  4 - n=  5$ & & 20 & 35& 17.16\%& 20& 30& 12.65\%\\ 
   4.65251$\mu$m & $n=  5 - n=  7$ & & 20 & 35& 14.92\%& 20 & 30& 10.89\%\\
   5.12726$\mu$m & $n=  6 - n=  10$ & & 20 & 30& 13.89\%& 20 & 25& 9.82\%\\
   5.90660$\mu$m & $n=  6 - n=  9$ & & 20 & 35& 15.93\%& 20 & 30& 11.62\%\\
   7.45782$\mu$m & $n=  5 - n=  6$ & & 20 & 40& 23.20\%& 20 & 30& 17.11\%\\
   7.50045$\mu$m & $n=  6 - n=  8$ & & 20 & 35& 20.22\%& 20&30& 14.66\%\\
   8.75768$\mu$m & $n=  7 - n=  10$ & & 20 & 35& 20.53\%& 20 &30& 14.83\%\\
   11.3056$\mu$m & $n=  7 - n=  9$ & & 20 & 40& 25.06\%& 20 & 30& 18.08\% \\
   12.3685$\mu$m & $n=  6 - n=  7$ & & 25 & 45& 28.31\%& 25 & 35& 20.94\%\\
   16.2047$\mu$m & $n=  8 - n=  10$ & & 25 & 45& 29.22\%& 25 & 35& 21.01\%\\
   19.0567$\mu$m & $n=  7 - n=  8$ & & 25 & 50& 32.43\%& 25 & 35& 24.16\%\\
   27.7958$\mu$m & $n=  8 - n=  9$ & & 25 & 50& 35.60\%& 25 &35& 26.80\%\\
\end{tabular}
\end{table*}

Our simple models demonstrate the dependence of 
the electron populations on the number of 
levels of the modeled atoms. For the pure 
hydrogen model we have tested the intensities 
of all recombination lines corresponding to the 
decay from $n=10$ to $n=2-8$, given in Table 
\ref{tab:linesh}. For the mixture of hydrogen 
and helium we monitor recombination lines up to 
$n=6$ in Table \ref{tab:lineshhe}. Many of 
these lines are used to obtain primordial 
helium abundances 
\citep{PorterFerland2007,Porter.R12Improved-He-I-emissivities-in-the-case-B-approximation,Porter.R13-CaseB-erratum}.

We can expect similar behaviour from diffuse 
gas emission in clouds illuminated by multi-
wavelength spectral energy distribution sources 
where the gas is almost completely ionized, 
i.e., H~II regions, Broad Line regions, and 
Planetary Nebulae. 

\section{Results: Dependence of the recombination lines intensities on the number of levels }
\label{sec:results}

To demonstrate how $\ell$-mixing influences 
electronic population, and, by extension, 
diffuse gas line emission, we increase the 
principal quantum number for which we computed 
resolved $\ell$-subshells until convergence in 
the line intensities. This will provide us with 
the number of resolved levels needed to be 
included in the calculations to obtain accurate 
line intensities. For our models, the minimum 
principal quantum number resolved in $\ell$ is 
$n=10$, which is the default for hydrogen in 
Cloudy \citep[see][]{C17}. Then, we increased 
the number of $\ell$-resolved $n$ in steps of 
five up to $n=70$. The rest of the levels 
remained collapsed up to $n=200$. We then 
calculated the minimum resolved $n$ necessary 
for converging the line intensity to 1\% and 5\%.

\subsection{Pure hydrogen gas}

Our first model used pure ionized hydrogen gas. We set the electron temperature of the 
plasma to 10000K, and its hydrogen density 
varied between $\n_H = 10^0 - 10^7\text{cm}
^{-3}$. The hydrogen density variation impacts 
the $\ell$-changing rates and can slightly 
affect critical densities in Figure 
\ref{f:ncritfitT4}. To study the influence of 
the number of resolved levels in Cloudy's 
predictions, we have progressively increased 
the highest resolved principal quantum number 
in 5 units from $n=10$ to $70$. Meanwhile, we 
kept the total number of $n$-shells at $n=200$. 
We have monitored the convergence of the 
intensities of all lines in Table 
\ref{tab:linesh}.

\begin{figure*}
    \centering
        \includegraphics[width=0.4\textwidth]{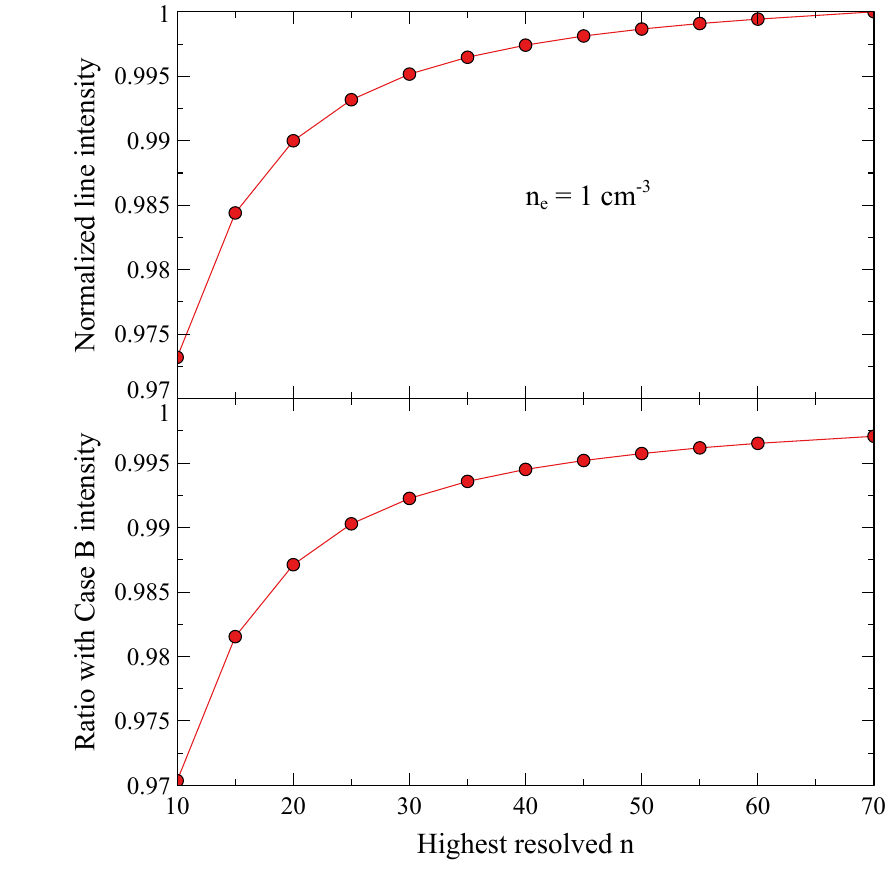}
\qquad
        \includegraphics[width=0.4\textwidth]{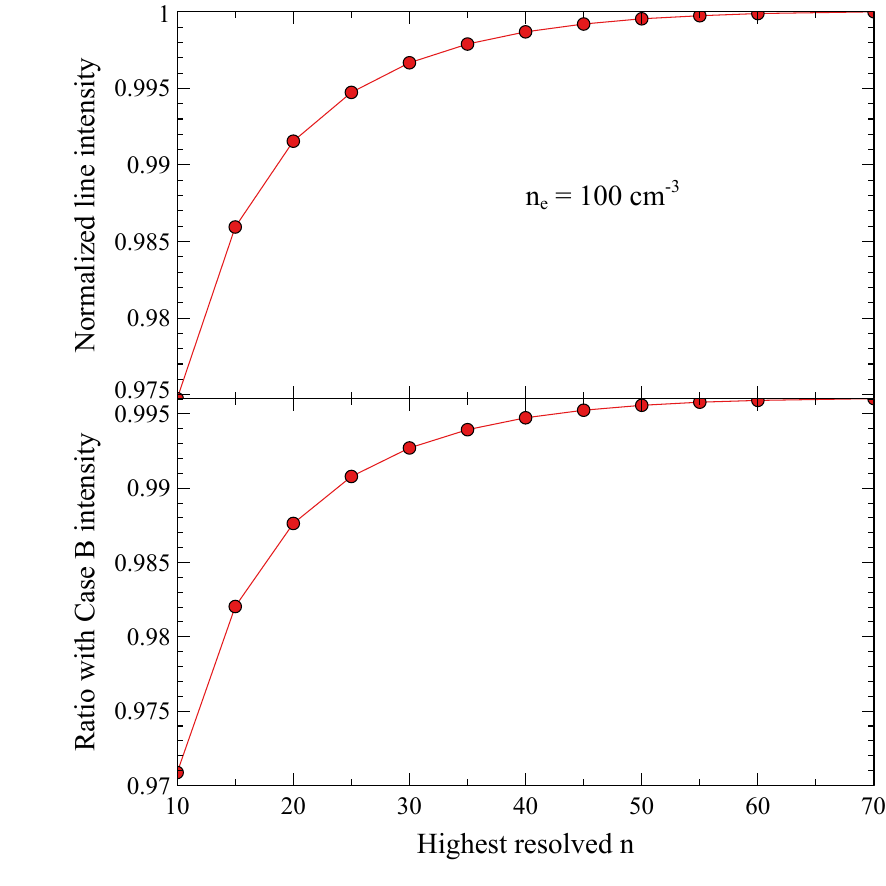}

        \includegraphics[width=0.4\textwidth]{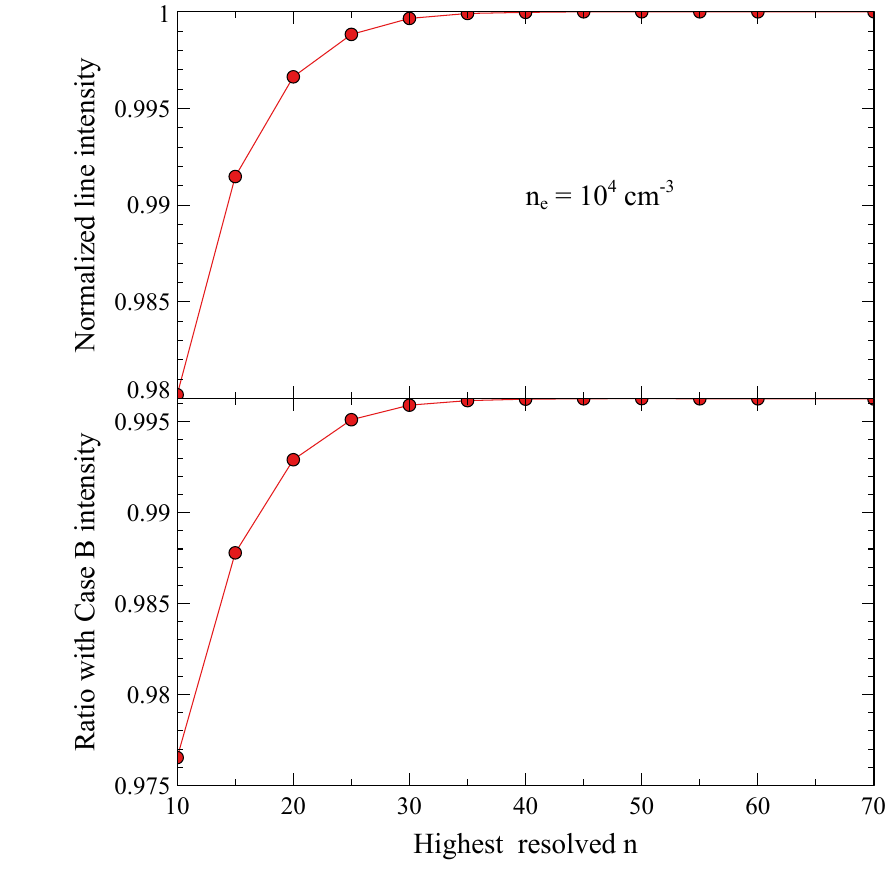}
\qquad
        \includegraphics[width=0.4\textwidth]{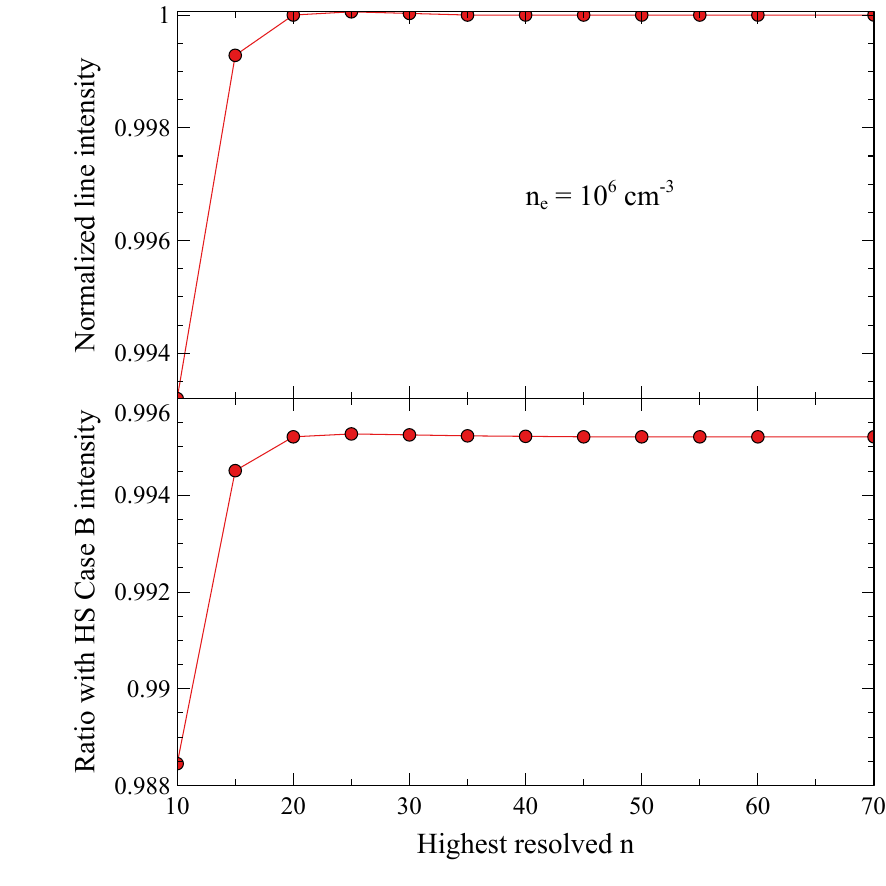}
    \caption{Normalized H$\alpha$ line intensity for a pure hydrogen gas at $T=10^4$K as a function of the highest resolved $n$ for different hydrogen densities. In each of the panels, the top graph represents the intensity normalized to the highest value obtained, and the bottom graph is the ratio with the case B intensity from \citet{Hummer1987}. Convergence at $<1$\% needs a  highest resolved level $n=20$ except for densities $\n_\text{H} \geq 10^6 \text{cm}^{-3}$.}
    \label{f:halpha}
\end{figure*}

\begin{figure}
    \centering
    \includegraphics[width=0.5\textwidth]{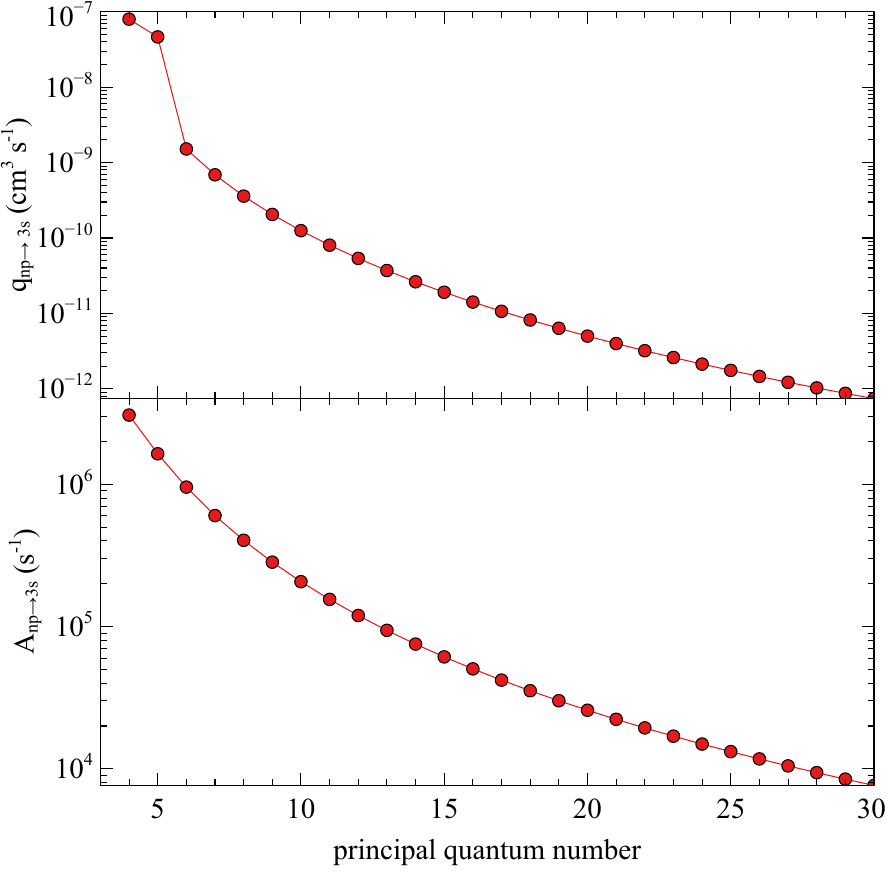}
    \caption{Top: Electron collision effective 
    coefficients for hydrogen $np\to 3s$ 
    transitions. Bottom: Radiative transition 
    probabilities (A-values) for $np\to 3s$.  }
    \label{f:transto3}
\end{figure}

In Figure \ref{f:halpha}, we show the effect 
that  insufficient resolved levels in the 
hydrogen atom produce on 
the H$\alpha$ line. In the Figure, we represent 
the predicted intensity as a function of the 
highest resolved principal quantum number 
considered in the calculation. For H$\alpha$, 
the critical density of the upper level ($n=3$) 
is $\n^\text{crit}\approx 10^9\text{cm}^{-3}$, 
well above the densities of most astrophysical 
nebulae, so Cloudy keeps the level $n=3$ 
resolved by default. However, electrons in 
upper $n$-shells influence the population of 
this level by decay. How the electron 
population of the $\ell$-subshells of these 
upper levels is distributed will depend mainly 
on the hydrogen density, as shown in Figures 
\ref{f:ncritfitT4} and \ref{f:ncritvsT}. It is 
important to include enough levels resolved in $
\ell$ to account for the correct electronic 
cascade towards these low levels. For example, 
the default calculation in Cloudy includes $
\ell$-resolved levels up to $n=10$, which will 
have a statistical electron population on the $
\ell$-subshells when the density is well above 
$\n_\text{H} = 10^5 \text{cm}^{-3}$ for 
$T=10^4$K (see Figure \ref{f:ncritfitT4}). In 
Figure \ref{f:halpha}, this is only happening 
at the bottom right panel, for densities 
$\n_\text{H} \geq 10^6 \text{cm}^{-3}$. At 
$\n_\text{H} = 10^4 \text{cm}^{-3}$, we need to 
increase the resolved levels to $n=20$ 
($\n^\text{crit} > 100 \text{cm}^{-3}$) to 
achieve convergence. In Figure \ref{f:halpha}, 
the highest resolved level needed for a 1\% 
converged prediction of the intensity of H$
\alpha$ does not change at lower densities: the 
transition probabilities from levels 
immediately over $n = 3$ dominate over the 
upper levels. As shown in Figure 
\ref{f:transto3}, radiative and collisional 
transition probabilities to $n=3$ decrease 
rapidly for $\Delta n> 1$, and levels over 
$n=15$ contribute only marginally to the 
electron population of the upper level. Even if 
levels with $n \geq 15$ are not adequately 
resolved, their electron population will 
preferentially go to immediate lower levels and 
will not directly contribute to the H$\alpha$ 
line. Note that the jump at $n=5$ on the 
effective collisional rates in Figure \ref{f:transto3} comes from 
the change from the R-matrix approach to the Born approximation 
(section \ref{sec:nchanging}) and is probably an artifice. 
However, suppose the coefficients for $n>5$ would be one order of 
magnitude higher and smoothly join with $n=5$. In that case, we 
can speculate that this would only slightly change Figure 
\ref{f:halpha}, as the effective coefficients for $n=15\to 3$ are 
already two orders of magnitude lower than $n=6\to 3$. Better 
collisional data is indeed necessary to obtain more accurate 
results.

Table \ref{tab:linesh} shows the minimum 
highest resolved $n$ for which the predicted 
line intensity converges with the highest 
resolved level to 5\% and 1\% for $\n_\text{H} = 
1 \text{cm}^{-3}$ and $\n_\text{H} = 10^4 
\text{cm}^{-3}$. We have also included the 
percentile difference of the line intensity 
prediction between our largest models (highest 
resolved level $n=70$) and the default (highest 
resolved level $n=10$) to indicate the 
sensitivity of the line to the number of 
resolved levels. 

\begin{figure}
    \centering
    \includegraphics[width=0.5\textwidth]{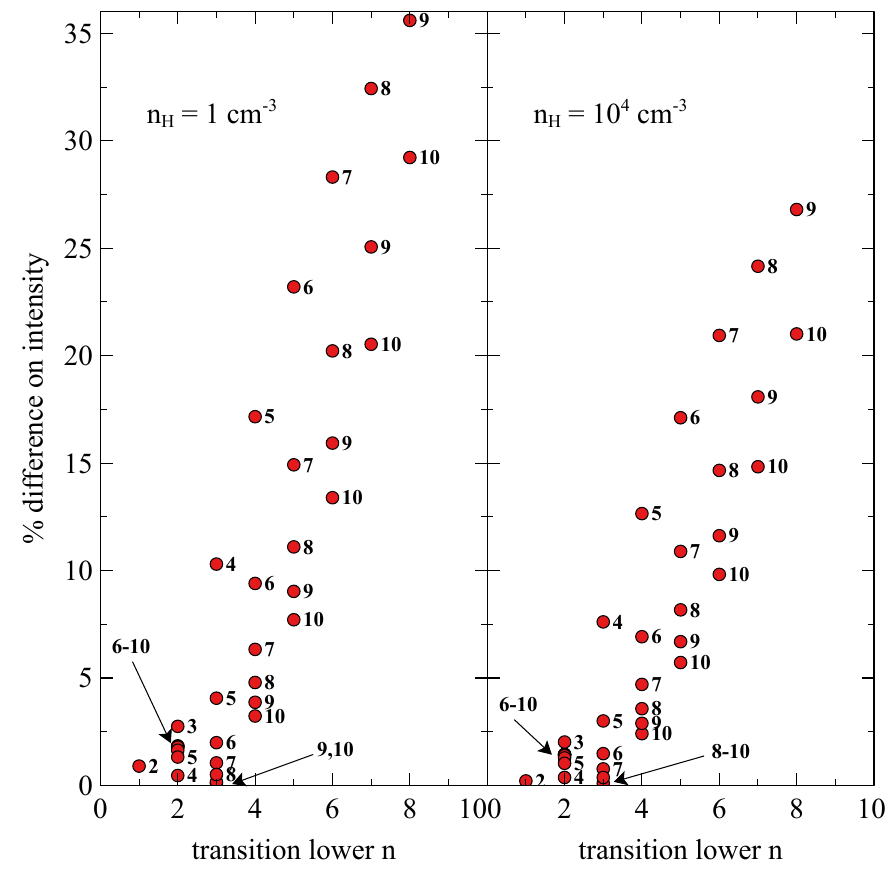}
    \caption{Percent difference for hydrogen 
    line intensity from calculations with 
    maximum resolved $n^\text{res}=70$ to one 
    with maximum resolved $n^\text{res}=10$ 
    (Cloudy default) as a function of the lower 
    $n$ of the transition. The tag numbers next 
    to each data point are the principal 
    quantum number $n$ for the upper level. }
    \label{f:linesvsdiff}
\end{figure}

\begin{figure}
    \centering
    \includegraphics[width=0.5\textwidth]{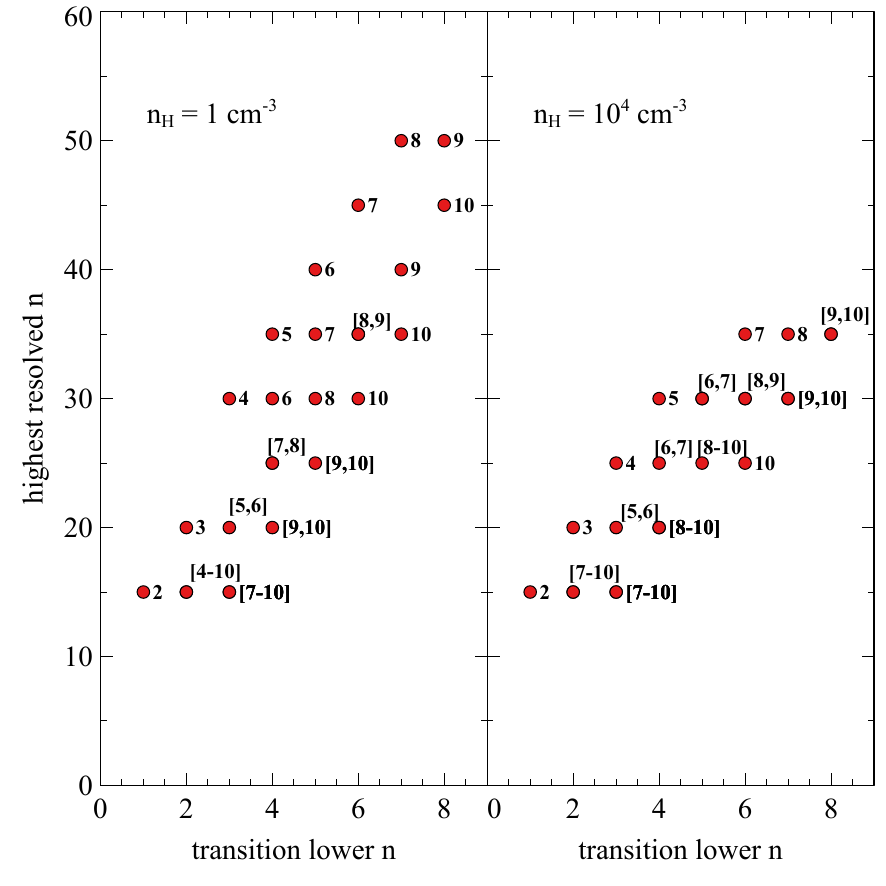}
    \caption{ Highest resolved principal 
    quantum number $n$ needed in our pure 
    hydrogen models to achieve convergence at 1\% 
    for the recombination line intensities as 
    a function of the lower $n$ of the 
    transition for different densities. The tags with lists of numbers or intervals next 
    to each data point are the principal 
    quantum number $n$ for the upper levels. }
    \label{f:linesvsres}
\end{figure}

The variation in the number of resolved levels 
does not strongly influence the upper lines of the 
Balmer series. Their better convergence 
can be understood considering the 
dominance of $\Delta n =1 $ radiative 
transitions, making the electron flux to $
\Delta n >1$ residual and the convergence 
errors smaller. This effect is more noticeable 
for the Paschen series, where the highest error 
is on the P$\alpha$ line (10.30\% for 
$\n_\text{H}= 1 \text{cm}^{-3}$ and 7.61\% for 
$\n_\text{H} = 10^4 \text{cm}^{-3}$). As seen in Figure 
\ref{f:transto3}, the electron effective 
coefficients and the radiative transition rates 
to $3s$ from different upper $n$'s
dramatically decay as $n$ increases. As we move 
towards the infrared to higher series, it takes 
more resolved levels to achieve convergence 
(Table \ref{tab:linesh}) because the upper 
levels of the transitions are closer to the highest resolved 
levels and more affected by electrons cascading 
from unresolved levels. These two effects, the 
predominance of lower $\Delta n$ decays and the 
larger deviation on the electronic populations 
of the upper levels closer to the resolved 
limit, combine to make the strongest lines from 
progressively higher series more sensitive to 
the number of resolved levels. That is shown in 
Figure \ref{f:linesvsdiff}, where we represent 
the percent differences in intensity plotted 
against the lowest level of the transition. In 
this Figure, we have tagged each data point 
with the upper level of the corresponding 
transition. Larger lower $n$'s tend to have a 
larger error, while larger upper levels that 
decay to the same lower level have less error. 
At higher densities, the critical densities of 
a smaller number of levels will be over the 
plasma density, so a lower number of resolved 
levels is required. That can be seen in Figure 
\ref{f:linesvsres}, where we represent the 
highest resolved level to achieve 1\% of 
convergence. The highest resolved $n$ needed 
tends to decrease for higher densities. It is, 
however, still around $n=35$ for NIR hydrogen 
transitions at electron densities of $\n_e=10^4 
\text{cm}^{-3}$.  

\subsection{Helium recombination lines }
\label{sec:herec}

A second set of models consisted of an ionized 
hydrogen and helium mixture gas where the 
helium abundance is 10\% with respect to 
hydrogen. The gas is illuminated with a 
monochromatic radiation of 2Ryd, enough to 
ionize hydrogen and also He atoms to He$^+$ in 
typical conditions for an H~II region. Hydrogen 
densities were the same as in the pure hydrogen 
gas case models. The default highest resolved 
level for the helium atom in a cloudy 
calculation is $n=6$ \citep{C17}, corresponding 
to a critical density of $\n^\text{crit} \approx 
1.7\times 10^8 \text{cm}^{-3}$. It is clear 
that in the range of densities considered in 
this study, where the highest hydrogen density 
$\n_\text{H} = 10^7 \text{cm}^{-3} $, levels 
higher than $n=6$ should be resolved. As we 
focus on the convergence of the emission lines 
of the recombined helium atoms, we have deemed 
it sufficient to start our models with the 
highest resolved level $n^\text{res} = 10$. As 
with the pure hydrogen gas, we have 
successively increased the number of resolved 
levels until $n=70$, where convergence was 
achieved for all the lines we considered. Note 
that this will not be necessarily the case for lines 
belonging to high-$n$ transitions in the FIR or 
the radio regions of the spectrum. We have 
considered transitions corresponding to 
electron decays between helium levels with 
principal quantum numbers $n=1$ to $n=6$, 
listed in Table \ref{tab:lineshhe}, covering a 
spectral range from the UV to the NIR. For 
simplicity, we kept only those lines that could 
be important for the determination of 
primordial helium abundances, as in 
\citet{PorterFerlandMacAdam2007}.

\begin{table*}[p]
    \centering
        \caption{ He I recombination lines 
        tested in our hydrogen and helium 
        mixture model at $T=10^4$K. The columns in the right contain the maximum 
        resolved principal quantum number $n$ 
        for which the line intensity converge 
        for less than 5\% and 1\% with respect 
        to the models with maximum resolved 
        levels $n^\text{res}= n-5$ (convergence ratios start to be calculated at $n=15$). We show these at two 
        relevant hydrogen densities of $\n_\text{H} = 
        1\text{cm}^{-3}$ (corresponding to the 
        interstellar medium) and $\n_\text{H} = 
        10^4\text{cm}^{-3}$ (H~II regions). For each density, the maximum percentage difference induced by changing the maximum resolved level from  $n = 70$ to $n = 10$ is also shown (columns "Diff.").}
    \label{tab:lineshhe}
    \begin{tabular}{c|c|c|c|c|c|c|c|c|c}
       transition type & $\lambda$ & transition & comments & \multicolumn{3}{|c|}{Convergence at $\n_\text{H}=1\text{cm}^{-3}$}  & \multicolumn{3}{|c|}{Convergence at $\n_\text{H}=10^4\text{cm}^{-3}$} \\
      &&&& n ($<$5\%) & n ($<$1\%) & Diff. &n ($<$5\%) & n ($<$1\%) & Diff\\
      \hline
      \hline 
    \multirow{19}{*}{singlet-singlet} & 512.099\AA & $1\,^1S -  6\,^1P_1$ & & 15 & 20& 3.11\%& 15 & 20 & 2.68\%\\
    & 515.617\AA & $1\,^1S -  5\,^1P_1$ & & 15 & 20& 3.59\%& 15 & 20 & 3.07\%\\
    & 522.213\AA & $1\,^1S -  4\,^1P_1$ & & 15 & 20& 3.21\%& 15 & 20 & 2.71\%\\
    & 537.030\AA & $1\,^1S -  3\,^1P_1$ & & 15 & 20& 2.39\%& 15 & 20 & 1.95\%\\
    & 584.334\AA & $1\,^1S -  2\,^1P_1$ & & 15 & 15& 1.26\%& 15 & 15 & 0.50\%\\
    & 3447.59\AA & $2\,^1S -  6\,^1P_1$ & & 15 & 20& 3.12\%& 15 & 20 & 2.72\% \\
    & 3964.73\AA & $2\,^1S -  4\,^1P_1$ & & 15 & 20& 3.22\%& 15 & 20 & 2.75\% \\
     & 4143.76\AA & $2\,^1P_1 -   6\,^1D_2$ & & 15 & 15& 0.86\%& 15 & 15 & 0.70\% \\
     & 4168.97\AA & $2\,^1P_1 -   6\,^1S$ & & 15 & 15& 0.60\%& 15 & 15 & 0.60\% \\
     & 4387.93\AA & $2\,^1P_1 -   5\,^1D_2$ && 15 & 15& 0.19\%& 15 & 15 & 0.13\%  \\
     & 4437.55\AA & $2\,^1P_1 -   5\,^1S$ & & 15 & 15& 0.71\%& 15 & 15 & 0.68\% \\
     & 4921.93\AA & $2\,^1P_1 -   4\,^1D_2$ & & 15 & 15& 1.41\%& 15 & 15 & 1.26\% \\
     & 5015.68\AA & $2\,^1S -  3\,^1P_1$ & & 15 & 20& 2.40\%& 15 & 20 & 1.99\% \\
     & 5047.74\AA & $2\,^1P_1 -   4\,^1S$ & & 15 & 15& 0.20\%& 15 & 15 & 0.23\%\\
     & 6678.15\AA & $2\,^1P_1 -   3\,^1D_2$ & & 15 & 25& 7.42\%& 15 & 25 & 5.55\%\\
     & 7281.35\AA & $2\,^1P_1 -   3\,^1S$ & & 15 & 15& 0.62\%& 15 & 15 & 0.36\% \\
     & 9603.44\AA & $3\,^1S -   6\,^1P_1$ & & 15 & 20& 3.12\%& 15 & 20 & 2.72\% \\
     & 12790.5\AA & $3\,^1D_2 -   5\,^1F_3$ & & 15 & 25& 8.08\%& 15 & 25 & 6.49\% \\
     & 20581.3\AA & $2\,^1S -   2\,^1P_1$ & & 15 & 15& 1.25\%& 15 & 15 & 0.49\% \\    
\hline
    \multirow{23}{*}{triplet-triplet} & 2829.08\AA & $2\,^3S -  6\,^3P$ & & 15 & 20& 3.16\%& 15 & 20 & 2.65\%  \\
     & 2945.10\AA & $2\,^3S -  5\,^3P$ & & 15 & 20& 3.99\%& 15 & 20 & 3.41\% \\
     & 3187.74\AA & $2\,^3S -  4\,^3P$ & & 15 & 20& 3.66\%& 15 & 20 & 3.00\% \\
     & 3819.62\AA & $2\,^3P_J -  6\,^3D$ & Blend of $J=0,1,2$  & 15 & 15& 0.14\%& 15 & 15 & 0.06\% \\
     & 3867.49\AA & $2\,^3P_J -  6\,^3S$ & Blend of $J=0,1,2$ & 15 & 25& 5.87\%& 15 & 25 & 5.44\% \\
     & 3888.64\AA & $2\,^3S -  3\,^3P$ & & 15 & 20& 2.94\%& 15 & 20 & 2.20\% \\
     & 4026.21\AA & $2\,^3P_J -  5\,^3D$ & Blend of $J=0,1,2$ & 15 & 15& 0.05\%& 15 & 15 & 0.14\% \\
     & 4120.84\AA & $2\,^3P_J -  5\,^3S$ & Blend of $J=0,1,2$ & 15 & 20& 1.86\%& 15 & 20 & 1.27\% \\
     & 4471.50\AA & $2\,^3P_J -  4\,^3D$ & Blend of $J=0,1,2$ & 15 & 15& 1.41\%& 15 & 15 & 1.23\% \\
     & 4713.17\AA & $2\,^3P_J -  4\,^3S$ & Blend of $J=0,1,2$ & 15 & 15& 1.01\% & 15 & 15 & 0.62\%\\
     & 5875.66\AA & $2\,^3P_J -  3\,^3D$ & Blend of $J=0,1,2$ & 15 & 25& 7.15\%& 15 & 25 & 5.28\% \\
     & 7065.25\AA & $2\,^3P_J -  3\,^3S$ & Blend of $J=0,1,2$ & 15 & 15& 0.46\%& 15 & 15 & 0.28\%  \\
    & 8361.73\AA & $3\,^3S -  6\,^3P$ & & 15 & 20& 3.15\%& 15 & 20 & 2.65\% \\
     & 9463.58\AA & $3\,^3S -   5\,^3P$ & & 15 & 20& 4.00\%& 15 & 20 & 3.41\% \\
     & 10830.2\AA & $2\,^3S -   2\,^3P_J$ & Blend of $J=0,1,2$ & 15 & 15& 0.38\%& 15 & 15 & 0.03\% \\
     & 10913.0\AA & $3\,^3D -   6\,^3F$ & & 15 & 25& 5.09\%& 15 & 20 & 4.43\% \\
     & 12527.5\AA & $3\,^3S -   4\,^3P$ & & 15 & 20& 3.67\%& 15 & 20 & 2.99\% \\
     & 12784.9\AA & $3\,^3D -   5\,^3F$ & & 15 & 25& 8.07\%& 15 & 25 & 6.67\% \\
     & 12846.0\AA & $3\,^3P -   5\,^3S$ & & 15 & 20& 1.86\%& 15 & 20 & 1.27\% \\
     & 12984.9\AA & $3\,^3D -   5\,^3P$ & & 15 & 20& 3.99\%& 15 & 20 & 3.41\% \\
     & 18685.4\AA & $3\,^3D -   4\,^3F$ && 20 & 35& 16.32\%& 20 & 30 & 13.10\%  \\
     & 19543.1\AA & $3\,^3D -   4\,^3P$ & & 15 & 20& 3.67\%& 15 & 20 & 3.00\% \\
    & 21120.2\AA & $3\,^3P -   4\,^3S$ && 15 & 15& 1.00\%& 15 & 15 & 0.62\%  \\
    \hline
        \multirow{3}{*}{singlet-triplet}& 591.409\AA & $1\,^1S -  2\,^3P_0$ & & 15 & 15& 0.38\%& 15 & 15 & 0.10\% \\
    & 625.563\AA & $1\,^1S -  2\,^3S$ & & 15 & 15& 0.12\%& 15 & 15 & 0.13\%\\
     & 8863.66\AA & $2\,^3S -  2\,^1P_1$ & & 15 & 15& 1.25\%& 15 & 15 & 0.49\% \\
\end{tabular}
\end{table*}

Table \ref{tab:lineshhe} lists the minimum 
resolved principal quantum number needed in our 
models to achieve a 5\% and 1\% convergence on 
the line intensity. Helium recombination lines 
in Table \ref{tab:lineshhe} present the same 
competing factors as in the case of pure 
hydrogen: the dominance of $\Delta n = 1$ and 
the closeness of the high levels to the lowest 
non-resolved level. However, in this case, two 
spin systems, singlet and triplet, and the 
split of the degeneracy of triplet levels 
complicate the picture. Moreover, as stated in 
section \ref{sec:critdens} and shown in Figure 
\ref{f:ncrithevsT}, critical densities do not 
follow the same pattern for different $n$ at 
different temperatures. As an illustration, in 
Table \ref{tab:selectlineshe}, we have taken 
the lines corresponding to spin-conserved 
transitions to levels $1\,^1S$ and $2\,^1S$. 
The differences between calculations with 
maximum resolved $n^\text{res}=10$ and 
$n^\text{res} =70$ increase up to $n=5$, to 
decrease again from there. In Table 
\ref{tab:selectlineshe}, the electron collision 
transition rates from the upper-level $n=6$ are 
at least an order of magnitude smaller than from smaller 
principal quantum numbers, revealing that $n=6$ 
is not as strongly connected with $n=2$ or 
$n=1$ as the other levels with smaller $n$, 
mitigating the influence of the electron 
populations from a wrongly statistically 
populated level ahead of it. This influence is 
dominant for the transitions with $n\leq 5$, 
which have an increasing error as the level is 
closer to the maximum resolved level. 
This effect is possibly artificial because the 
separation of collisional rates between $n=5$ 
and $n=6$ coincides with different sets of 
collisional data obtained from different 
theoretical approaches (see section 
\ref{sec:nchanging}). As in the case of hydrogen, better collisional data is necessary 
to assess this problem. 

A final exception comes from the transitions 
from upper $n\ell$-shells with larger angular 
momentum. These transitions show a high percent 
difference for different maximum resolved 
levels. This is explained by the more effective 
decay of the yrast levels through $\Delta n =1$ 
transitions. Thus, they are more influenced by 
the forced statistical population of the upper 
non-resolved levels. This is the case of the 
lines with the more significant uncertainty on 
Table \ref{tab:lineshhe}. For example,  the 
transition  ($2\,^1P_1 - 3\,^1D_2$) $\lambda 
6678.15$\AA ~has a maximum percent difference 
of 7.42\%, while  ($2\,^1P_1 - 3\,^1S$) $
\lambda7281.35$\AA ~only amounts to 0.62\%. We 
obtain similar comparative ratios for other 
transitions with a high uncertainty, like  
($2\,^3P_J -   3\,^3D$) $\lambda 5875$\AA 
~(blended in $J$) compared to ($2\,^3P_J -   
3\,^3S$) $\lambda 7065.25$\AA, or  ($3\,^3D -   
4\,^3F$) $\lambda 18685.1$\AA ~compared to 
$3\,^3D -   4\,^3P$ $\lambda 19543.1$\AA 

\begin{table}
\centering
\caption{Singlet-singlet transitions to $1\,^1S$ 
and $2\,^1S$ from Table \ref{tab:lineshhe} for 
an electron temperature of $T = 10^4$K. 
The third columns lists the converging difference between $n^\text{res}=70$ and $n^\text{res} =10$ for electron density $\n_e = 1 \text{cm}^{-3}$ as in Tables 
\ref{tab:linesh} and \ref{tab:lineshhe}. The 
last two columns correspond to the collisional 
and radiative decay rates of the transitions.}
\label{tab:selectlineshe}
    \begin{tabular}{c|c|c|c|c}
    $\lambda$ & transition &  Diff & $q_{n^\prime\,^1P\to n\,^1S}$ & $A_{n^\prime\,^1P\to n\,^1S}$\\
    &&&&\\
    (\AA) & & & ($\text{cm}^3\text{s}^{-1}$) & ($\text{s}^{-1}$)\\
      \hline
      \hline 
    512.099 & $1\,^1S -  6\,^1P_1$ & 3.11\% & $2.26\times10^{-12}$ & $7.32\times10^7$\\
    515.617 & $1\,^1S -  5\,^1P_1$ & 3.59\% & $2.64\times10^{-11}$ & $1.26\times10^8$\\
    522.213 & $1\,^1S -  4\,^1P_1$ & 3.21\% & $4.33\times10^{-11}$ & $2.43\times10^8$\\
    537.030 & $1\,^1S -  3\,^1P_1$ & 2.39\% &$1.06\times10^{-10}$ & $5.66\times10^8$\\
    584.334 & $1\,^1S -  2\,^1P_1$ & 1.26\% & $3.74\times10^{-10}$ & $1.80\times10^9$\\
    \hline
    3447.59 & $2\,^1S -  6\,^1P_1$ & 3.12\% & $5.45\times10^{-11}$ & $2.27\times10^6$\\
    3964.73 & $2\,^1S -  4\,^1P_1$ & 3.22\% & $3.27\times10^{-9}$ & $6.95\times10^6$\\
    5015.68 & $2\,^1S -  3\,^1P_1$ & 2.40\% & $9.81\times10^{-9}$ & $1.34\times10^7$ \\
    20581.3 & $2\,^1S -   2\,^1P_1$ & 1.25\% & $5.54\times10^{-9}$ & $1.97\times10^6$\\  
    \end{tabular}
    \end{table}

\begin{figure}
    \centering
    \includegraphics[width=0.5\textwidth]{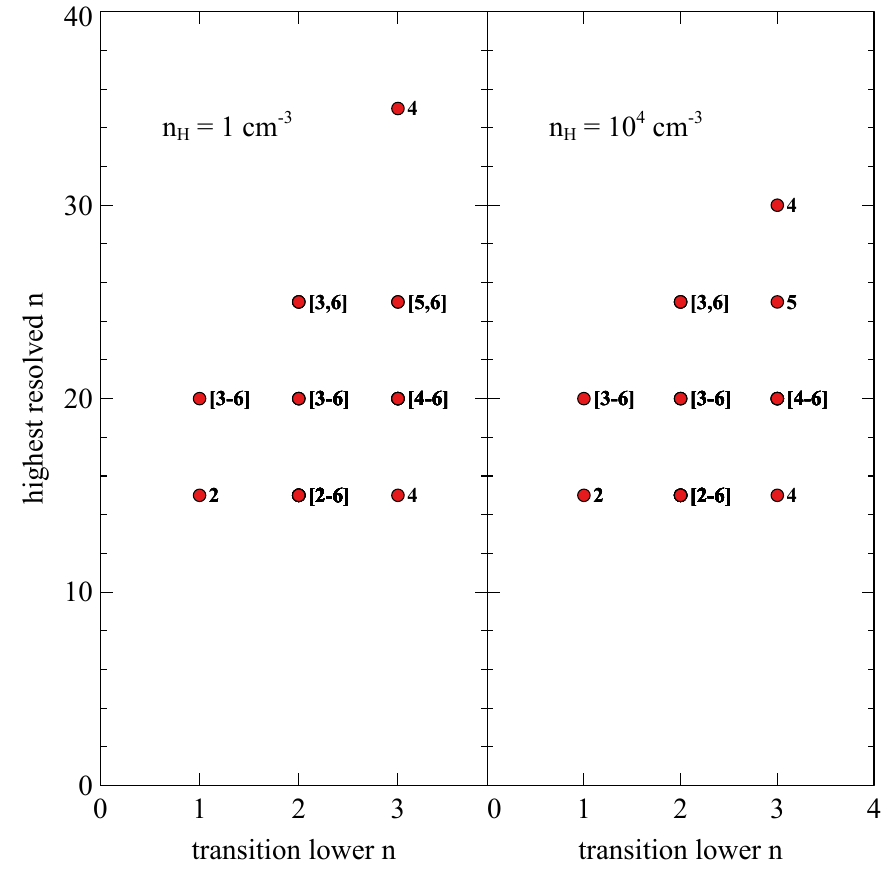}
    \caption{ Highest resolved principal 
    quantum number $n$ needed in our helium 
    atom model to achieve convergence for the 
    recombination lines at 1\% as a function of 
    the lower $n$ of the transition for 
    different densities. Note that each point 
    can represent multiple transitions. The tags with lists of numbers or intervals next 
    to each data point are the principal 
    quantum number $n$ for the upper levels.}
    \label{f:helinesvsres}
\end{figure}

While helium recombination lines of interest 
typically involve low levels, specifically in 
the optical range, their predicted intensities 
can still be significantly affected by the 
wrong choice of the number of resolved levels. 
This effect is more considerable when the hydrogen 
densities are smaller, so more levels fall 
under their critical densities. As shown in 
Figure \ref{f:helinesvsres}, convergence to 1\% 
needs at least a $n^\text{res} = 15$ for 
most optical lines, but it can get to 
$n^\text{res} = 35$ for infrared lines.  

\begin{figure*}
    \centering
        \includegraphics[width=0.45\textwidth]{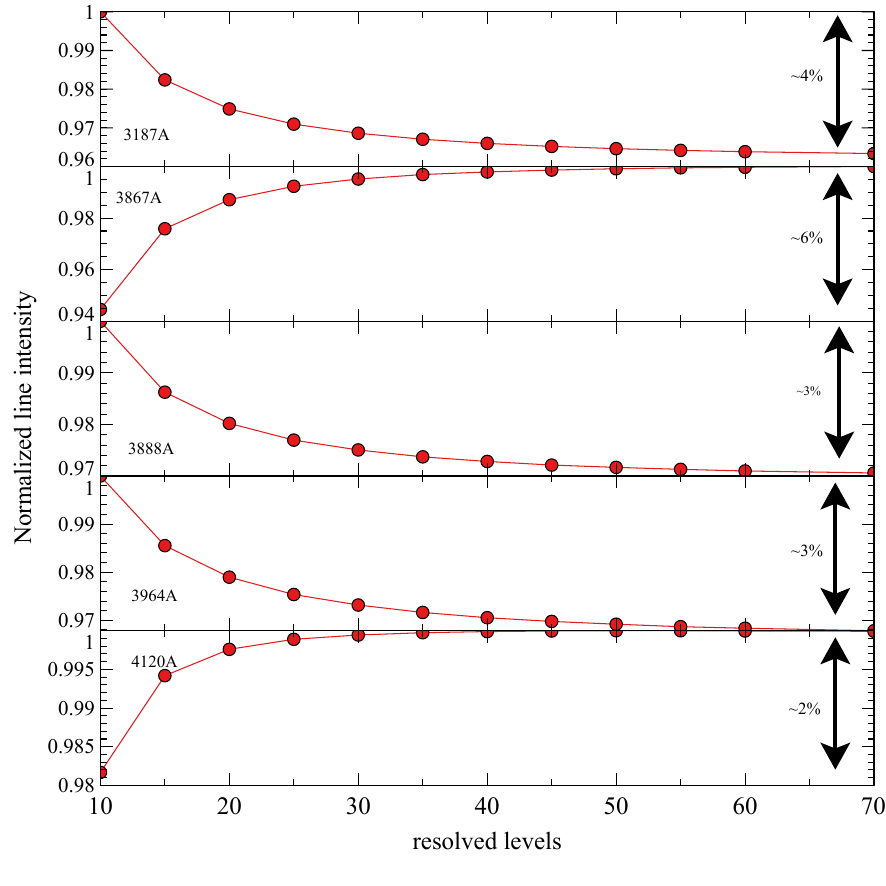}
\quad
        \includegraphics[width=0.45\textwidth]{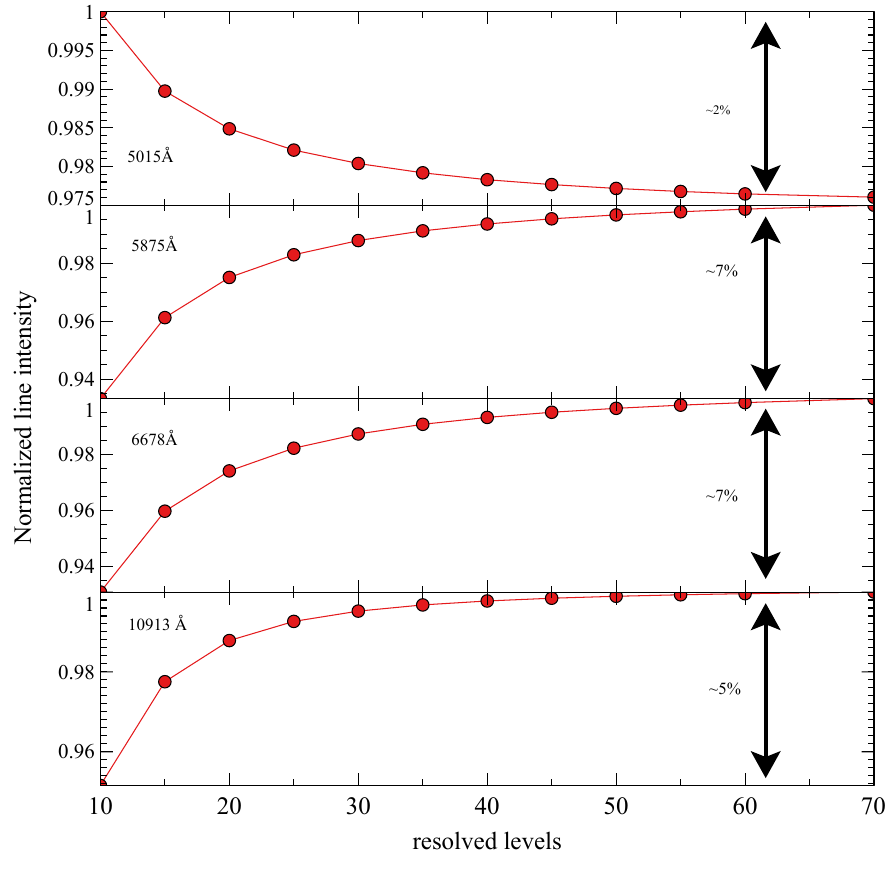}
\quad
        \includegraphics[width=0.45\textwidth]{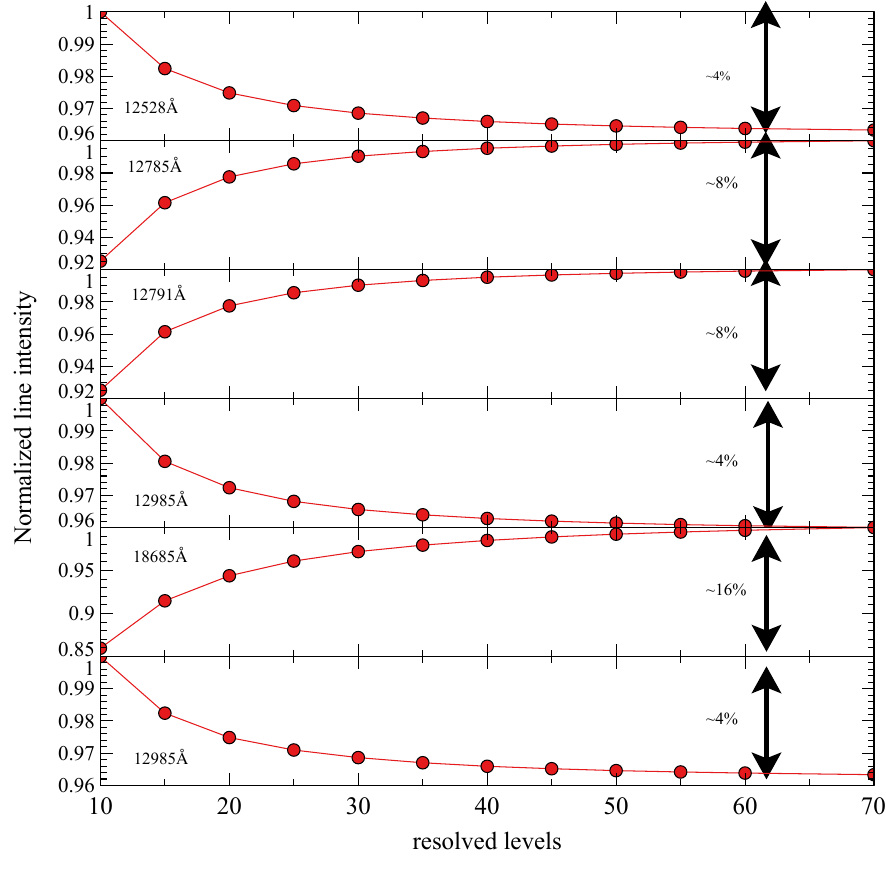}

    \caption{He~I recombination line normalized 
    line intensities for selected lines in a 
    hydrogen-helium gas, at $T=10^4$K and 
    density $\n_\text{H} = 1 \text{cm}^{-3}$, as 
    a function of the maximum resolved $n$ 
    included in the models.}
    \label{f:helines}
\end{figure*}

In Figure \ref{f:helines}, we show the 
convergence of the helium line intensities with 
the most significant percent differences for 
densities of $\n_\text{H} = 1 \text{cm}^{-3}$, 
corresponding to typical interstellar medium 
densities. Even when 1\% convergence is 
achieved fast at $n^\text{res} = 15-20$, except 
for some extreme cases, most of the lines keep 
slowly varying their intensity up to 
$n^\text{res} = 60-70$. As a rule of thumb, a 
safe value for the lines would be to set 
$n^\text{res} \approx 40-50$, well over the 
minimum value required for the critical density 
shown in Figure \ref{f:ncrithe}. This must be 
taken with care, and some lines might need a 
more conservative approach even at much higher 
densities. For example, ($3\,^3D -   4\,^3F$) 
$\lambda 18685.1$\AA ~has a 1\% convergence of 
the intensity at a maximum resoved level of 
$n^\text{res} = 30$ at $\n_\text{H} = 10^4 
\text{cm}^{-3}$, while Figure \ref{f:ncrithe} 
suggests a $n^\text{res} = 15-20$, well below 
that value.

\section{Conclusions}
\label{sec:discussion}

The main aim of this study is to predict the 
number of $n\ell$-resolved levels needed to 
obtain accurate line intensities from hydrogen 
and helium in photoionization models, which 
applies to the determination of primordial helium 
abundances from line emission on metal-poor 
galaxies. While we do not think modeling should 
be added to the error budget of primordial helium 
determination in table 1 of \citet{Peimbert2007}, 
it is crucial to consider atoms with an adequate 
number of levels to avoid the derived 
uncertainties on the line emissivities. This 
work will help quantify the number of levels to 
remove the uncertainties of high precision 
problems. 

In this paper, we have shown the dependence of 
the calculated hydrogen and helium recombination 
line intensities on the number of resolved 
$n\ell$-shells included in the photoionization 
models. Our results show that, especially at low 
densities, a high number of $n\ell$ resolved 
levels need to be included to get an accurate 
value of the line intensities. This error can be 
more significant for infrared lines with a high 
upper $n$. The temperature dependence on the 
critical densities emanating from the $\ell$-
changing cross-sections can be more significant 
for low level transitions inside helium atoms 
(Figure \ref{f:ncrithevsT}).

Our literature search found that different 
studies have varied ways of dealing with this 
problem. \citet{Storey1995} used the radiative to collisional excitations ratio to obtain the maximum resolved $n$ for which the transition $np\to n(n-1)$ has a 90\% chance to happen before a radiative transfer can occur. Using this method they find differences in 1\% in the departure coefficients of high $nl$-levels with respect to a fully resolved calculation. Some authors avoided the problem by accounting 
for a number of resolved levels in their 
photoionization models which is big enough to be over any 
considered critical densities, both in hydrogen 
and helium. For example, \citet{Porter2005} use 
$n^\text{res} = 100$ for modeling He~I emission 
in H~II regions. \citet{PorterFerlandMacAdam2007} 
and \citet{Porter.R09Uncertainties-in-theoretical-HeI-emissivities:-HII-regions} reduce the 
resolved levels to $n^\text{res} = 40$ and 
$n\leq100$, safely converging the helium lines of 
Table \ref{tab:lineshhe} for densities of 
$\n_\text{H} =10^4 \text{cm}^{-3}$. 
\citet{Porter.R12Improved-He-I-emissivities-in-the-case-B-approximation,Porter.R13-CaseB-erratum} set 
$n^\text{res}\leq100$ for a $n\leq101$ model for 
helium. While a calculation of this size will 
give accurate results for visible and most 
infrared lines, we deem it unnecessary to resolve 
as many levels for the densities considered in 
their study, $10^1\text{cm}^{-3} \leq \n_e \leq 
10^{14} \text{cm}^{-3}$, that are well over the 
critical densities for $n=100$. In their models 
for helium emission, \citet{DelZanna2022} use 
$n\leq100$ and $n^\text{res}\leq40$, which is 
suitable for all helium visible lines at most 
densities of nebular astrophysics. 
\citet{DelZanna2020} use the same, $n\leq100$ and 
$n^\text{res}\leq40$, for their models of the 
solar corona at electron densities of $\n_e = 
10^8\text{cm}^{-3}$ and temperatures of 
$T=10^6$K, corresponding to the critical 
densities of $n\approx5$ (Figs. \ref{f:ncrithe} 
and \ref{f:ncrithevsT}). Therefore, we can 
consider their solar and nebular results well 
converged. Recently, \citet{Balser2024} included 
up to $n^\text{res}=25$ for hydrogen and 
$n^\text{res}=20$ for helium in their simulations 
of the H~II region's emission lines for densities 
that ranged from $\n_e=10^1\text{cm}^{-3}$ to 
$\n_e=10^4\text{cm}^{-3}$. According to Tables 
\ref{tab:linesh} and \ref{tab:lineshhe}, this 
provides convergence of most of the optical 
recombination lines of hydrogen and helium. Other 
authors have used these results: \citet{Aver2015} 
worked with the emissivities provided by 
\citet{Porter.R12Improved-He-I-emissivities-in-the-case-B-approximation} on their study to 
obtain helium primordial abundances. Similarly 
\citet{Izotov2014} use the emissivities from the 
models from \citet{Porter.R09Uncertainties-in-theoretical-HeI-emissivities:-HII-regions}.

However, many studies that cite the latest papers 
of Cloudy \citep{C23,C17} omit the number of 
resolved levels accounted for in their models. It 
is important to note that if the commands that 
set the number of resolved and collapsed levels,

\begin{verbatim}
    database h-like hydrogen resolved levels 80
    database h-like hydrogen collapsed levels 70
    database he-like helium resolved levels 50
    database he-like helium collapsed levels 100,
\end{verbatim}

\noindent are ignored in the inputs, the default 
number of resolved levels is $n^\text{res}=10$ 
for hydrogen atoms and $n^\text{res}=6$ for 
helium. These numbers are insufficient at the 
densities and temperatures of most astrophysical 
plasmas.

Supported by the results of this work, we 
strongly recommend selecting a maximum number of 
resolved levels that will provide converged 
emissivities for the lines under interest. The 
most straightforward method is using Figures 
\ref{f:ncritfitT4}, \ref{f:ncritvsT}, 
\ref{f:ncrithe}, and \ref{f:ncrithevsT} to select 
the level at which the density is over the 
critical density and add at least ten principal 
quantum numbers to determine the maximum resolved 
level. We provide ASCII files with copies of 
Tables \ref{tab:linesh} and \ref{tab:lineshhe} 
attached to this paper for all densities 
($1\text{cm}^{-3} \leq \n_\text{H} \leq 10^7 
\text{cm}^{-3}$) considered in this work. These 
will help to provide a guide for the required 
convergence of the emissivites of specific lines.

\section*{Acknowledgements}

MC and GJF acknowledge support from NSF (1910687),
and NASA (19-ATP19-0188, 22-ADAP22-0139).
GJF acknowledges support from JWST through AR6428, AR6419, GO5354, and GO5018.

\section*{Data Availability}

The files attached with this paper include a version of Tables \ref{tab:linesh} and \ref{tab:lineshhe} for all densities considered: $1\text{cm}^{-3} \leq \n_\text{H} \leq 10^7$. These tables have been generated using the developed version of Cloudy C23.01 with versions of the following input script:

\begin{verbatim}
set save prefix "h40_4"
#
init file "honly.ini"
#
laser 2 ryd
ionization parameter -2
#
hden 4
constant temperature 4 log
set dr 0
#
case B 
#
database h-like hydrogen levels resolved 40
database h-like hydrogen levels collapsed 160
#
stop zone 1
iterate 2
#
print critical densities h-like
print lines column
#
save line list ".list" "linesh.dat" column absolute last no hash
save line list ratios ".rat" "linesh_ratio.dat" column last no hash
#
\end{verbatim}

Here, the number of resolved levels for hydrogen, density (\verb+hden+), and temperature can be changed to a grid. Input scripts for helium are similar.



\bibliographystyle{mnras}
\bibliography{bibliography2} 




\appendix

\section{Convergence at densities $1\text{\lowercase{cm}}^{-3} \leq \lowercase{\n}_\text{H} \leq 10^7\text{\lowercase{cm}}^{-3}$.}

Below, we reproduce Tables \ref{tab:linesh} and \ref{tab:lineshhe} for all other densities between $1\text{cm}^{-3} \leq \n_\text{H} \leq 10^7\text{cm}^{-3}$, corresponding the density range investigated in this work. These tables can serve as a guide for setting the number of levels to obtain accurate emissivities for He~I and H~I recombination lines from UV to IR.

\begin{table*}
    \centering
    \begin{sideways}
  \begin{minipage}{18cm}
        \caption{ H I recombination lines corresponding to the decays of the first eight series (starting with Balmer, $n=2$) up to $n=10$ that have been tested in our pure hydrogen plasma model at $T=10^4$K. The columns in the right contain the maximum resolved principal quantum number $n$ at which the line intensity converge for less than 5\% and 1\% with respect to the models with maximum resolved levels $n^\text{res} = n-5$ (convergence ratios start to be calculated at $n=15$). We show these at hydrogen densities of $10\text{cm}^{-3} \leq \n_\text{H} < 10^4$. For each density, the maximum percentage difference induced by changing the maximum resolved level from  $n = 70$ to $n = 10$ is also shown (columns "Diff.").}
    \label{tab:lineshlowdens}
    \begin{tabular}{c|c|c|c|c|c|c|c|c|c|c|c|}
       $\lambda$ & transition & comments & \multicolumn{3}{|c|}{Convergence at $\n_\text{H}=10\text{cm}^{-3}$} & \multicolumn{3}{|c|}{Convergence at $\n_\text{H}=10^2\text{cm}^{-3}$}& \multicolumn{3}{|c|}{Convergence at $\n_\text{H}=10^3\text{cm}^{-3}$}  \\
      &&& n ($<$5\%) & n ($<$1\%) & Diff. & n ($<$5\%) & n ($<$1\%)& Diff. &n ($<$5\%) & n ($<$1\%) & Diff.  \\
      \hline
      \hline
  1215.67\AA & $1\,^2S -   2\,^2P$ & Ly$\alpha$ & 15 & 15& 0.88\%& 15 & 15&0.82\% & 15 & 15&0.63\%\\
   3797.90\AA & $2\,^2S - n=  10$ & &15 & 15& 1.80\%& 15& 15& 1.74\% & 15& 15& 1.62\%\\ 
   3835.38\AA & $2\,^2S - n=  9$ & &15 & 15& 1.82\% & 15&15 & 1.77\%& 15&15 & 1.65\%\\
   3889.05\AA & $2\,^2S - n=  8$ & &15 & 15& 1.81\%& 15& 15& 1.76\%& 15& 15& 1.64\%\\
   3970.07\AA & $2\,^2S - n=  7$ & &15 & 15& 1.75\%& 15& 15 & 1.70\%& 15& 15 & 1.59\%\\
   4101.73\AA & $2\,^2S - n=  6$ & &15 & 15& 1.63\%& 15& 15 & 1.57\%& 15& 15 & 1.47\%\\ 
   4340.46\AA & $2\,^2S - n=  5$ & &15 & 15& 1.31\%& 15& 15& 1.27\%& 15& 15 & 1.18\%\\
   4861.32\AA & $2\,^2S - n=  4$ & H$\beta$ & 15 & 15& 0.45\%& 15& 15& 0.45\%& 15& 15 & 0.41\%\\
   6562.80\AA & $2\,^2S - n=  3$ & H$\alpha$ & 15 & 20 & 2.71\%& 15 & 20& 2.59\%& 15& 20 & 2.38\%\\
   9014.91\AA & $n=  3 - n=  10$ & & 15 & 15& 0.11\%& 15& 15& 0.11\% & 15& 15 & 0.11\%\\ 
   9229.02\AA & $n=  3 - n=  9$ & & 15 & 15& 0.15\%& 15& 15& 0.14\% & 15& 15 & 0.13\%\\ 
   9545.97\AA & $n=  3 - n=  8$ & & 15 & 15& 0.50\%& 15& 15& 0.48\%& 15& 15 & 0.44\%\\ 
   1.00494$\mu$m & $n=  3 - n=  7$ & & 15 & 15& 1.03\%& 15& 15& 0.98\%& 15& 15 & 0.91\%\\ 
   1.09381$\mu$m & $n=  3 - n=  6$ && 15 & 20 & 1.96\%& 15& 20& 1.88\%& 15& 20 & 1.74\%\\ 
   1.28181$\mu$m & $n=  3 - n=  5$ && 15 & 20& 3.93\%& 15& 20& 3.89\% & 15& 20 & 3.53\%\\
   1.73621$\mu$m & $n=  4 - n=  10$ & & 15 & 20& 3.17\%& 15 & 20& 3.05\% & 15& 20 & 2.82\%\\
   1.81741$\mu$m & $n=  4 - n=  9$ & & 15 & 20& 3.81\%& 15& 20& 3.66\%& 15& 20 & 3.39\%\\  
   1.87510$\mu$m & $n=  3 - n=  4$ & & 15 & 30& 10.15\%& 15& 30& 9.73\% & 15& 30 & 8.95\%\\
   1.94456$\mu$m & $n=  4 - n=  8$ & & 15 & 25& 4.71\%& 15& 25& 4.53\%& 15& 20 & 4.19\%\\  
   2.16553$\mu$m & $n=  4 - n=  7$ & & 15 & 25& 6.23\%& 15& 25& 5.98\%& 15& 25 & 5.53\%\\ 
   2.62515$\mu$m & $n=  4 - n=  6$ & & 15 & 30& 9.24\%& 15& 30& 8.86\%& 15& 25& 8.16\%\\
   3.03837$\mu$m & $n=  5 - n=  10$ & & 15 & 25& 7.58\%& 15 & 25& 7.28\%& 15& 25& 6.73\%\\
   3.29609$\mu$m & $n=  5 - n=  9$ & & 15 & 25& 8.88\%& 15 & 25& 8.53\%& 15& 25& 7.87\%\\
   3.73954$\mu$m & $n=  5 - n=  8$ & & 20 & 30& 10.91\%& 20 & 30& 10.46\%& 20& 30& 9.64\%\\
   4.05115$\mu$m & $n=  4 - n=  5$ & & 20 & 35& 16.92\%& 20& 35& 16.22\%& 20& 35& 14.89\%\\ 
   4.65251$\mu$m & $n=  5 - n=  7$ & & 20 & 35& 14.65\%& 20 & 30& 14.01\%& 20& 30& 12.87\%\\
   5.12726$\mu$m & $n=  6 - n=  10$ & & 20 & 30& 13.39\%& 20 & 30& 13.17\%& 20& 30& 11.60\%\\
   5.90660$\mu$m & $n=  6 - n=  9$ & & 20 & 35& 15.65\%& 20 & 35& 14.97\%& 20& 30& 13.74\%\\
   7.45782$\mu$m & $n=  5 - n=  6$ & & 20 & 40& 22.89\%& 20 & 40& 21.95\%& 20& 35& 20.12\%\\
   7.50045$\mu$m & $n=  6 - n=  8$ & & 20 & 35& 19.83\%& 20&35& 18.94\%& 20& 35& 17.34\%\\
   8.75768$\mu$m & $n=  7 - n=  10$ & & 20 & 35& 20.12\%& 20 &35& 19.21\%& 20& 35& 17.57\%\\
   11.3056$\mu$m & $n=  7 - n=  9$ & & 20 & 40& 24.56\%& 20 & 40& 23.42\%& 20& 35& 21.39\% \\
   12.3685$\mu$m & $n=  6 - n=  7$ & & 25 & 45& 27.98\%& 25 & 45& 26.85\%& 25& 40& 24.61\%\\
   16.2047$\mu$m & $n=  8 - n=  10$ & & 25 & 45& 28.61\%& 25 & 40& 27.26\%& 25& 40& 24.85\%\\
   19.0567$\mu$m & $n=  7 - n=  8$ & & 25 & 50& 32.14\%& 25 & 45& 30.90\%& 25& 40& 28.32\%\\
   27.7958$\mu$m & $n=  8 - n=  9$ & & 25 & 50& 35.41\%& 25 &50& 34.13\%& 25& 40& 31.32\%\\
\end{tabular}
\end{minipage}
 \end{sideways}
\end{table*}

\begin{table*}
    \centering
    \begin{sideways}
  \begin{minipage}{18cm}
        \caption{ H I recombination lines corresponding to the decays of the first eight series (starting with Balmer, $n=2$) up to $n=10$ that have been tested in our pure hydrogen plasma model at $T=10^4$K. The columns in the right contain the maximum resolved principal quantum number $n$ at which the line intensity converge for less than 5\% and 1\% with respect to the models with maximum resolved levels $n^\text{res} = n-5$ (convergence ratios start to be calculated at $n=15$). We show these at hydrogen densities of $10^5\text{cm}^{-3} \leq \n_\text{H} \leq 10^7$. For each density, the maximum percentage difference induced by changing the maximum resolved level from  $n = 70$ to $n = 10$ is also shown (columns "Diff.").}
    \label{tab:lineshhighdens}
    \begin{tabular}{c|c|c|c|c|c|c|c|c|c|c|c|}
       $\lambda$ & transition & comments & \multicolumn{3}{|c|}{Convergence at $\n_\text{H}=10^5\text{cm}^{-3}$} & \multicolumn{3}{|c|}{Convergence at $\n_\text{H}=10^6\text{cm}^{-3}$}& \multicolumn{3}{|c|}{Convergence at $\n_\text{H}=10^7\text{cm}^{-3}$}  \\
      &&& n ($<$5\%) & n ($<$1\%) & Diff. & n ($<$5\%) & n ($<$1\%)& Diff. &n ($<$5\%) & n ($<$1\%) & Diff.  \\
      \hline
      \hline
  1215.67\AA & $1\,^2S -   2\,^2P$ & Ly$\alpha$ & 15 & 15& 0.01\%& 15 & 15&0.02\% & 15 & 15&0.02\%\\
   3797.90\AA & $2\,^2S - n=  10$ & &15 & 15& 1.07\%& 15& 15& 0.26\% & 15& 15& 0.60\%\\ 
   3835.38\AA & $2\,^2S - n=  9$ & &15 & 15& 1.11\% & 15&15 & 0.44\%& 15&15 & 0.24\%\\
   3889.05\AA & $2\,^2S - n=  8$ & &15 & 15& 1.11\%& 15& 15& 0.51\%& 15& 15& 0.04\%\\
   3970.07\AA & $2\,^2S - n=  7$ & &15 & 15& 1.06\%& 15& 15 & 0.52\%& 15& 15 & 0.05\%\\
   4101.73\AA & $2\,^2S - n=  6$ & &15 & 15& 0.97\%& 15& 15 & 0.48\%& 15& 15 & 0.08\%\\ 
   4340.46\AA & $2\,^2S - n=  5$ & &15 & 15& 0.78\%& 15& 15& 0.39\%& 15& 15 & 0.08\%\\
   4861.32\AA & $2\,^2S - n=  4$ & H$\beta$ & 15 & 15& 0.30\%& 15& 15& 0.15\%& 15& 15 & 0.03\%\\
   6562.80\AA & $2\,^2S - n=  3$ & H$\alpha$ & 15 & 20 & 1.44\%& 15 & 15& 0.68\%& 15& 15 & 0.13\%\\
   9014.91\AA & $n=  3 - n=  10$ & & 15 & 15& 0.02\%& 15& 15& 0.44\% & 15& 15 & 0.75\%\\ 
   9229.02\AA & $n=  3 - n=  9$ & & 15 & 15& 0.06\%& 15& 15& 0.24\% & 15& 15 & 0.71\%\\ 
   9545.97\AA & $n=  3 - n=  8$ & & 15 & 15& 0.23\%& 15& 15& 0.19\%& 15& 15 & 0.23\%\\ 
   1.00494$\mu$m & $n=  3 - n=  7$ & & 15 & 15& 0.50\%& 15& 15& 0.26\%& 15& 15 & 0.16\%\\ 
   1.09381$\mu$m & $n=  3 - n=  6$ && 15 & 15 & 1.00\%& 15& 15& 0.48\%& 15& 15 & 0.16\%\\ 
   1.28181$\mu$m & $n=  3 - n=  5$ && 15 & 20& 2.09\%& 15& 15& 0.99\% & 15& 15 & 0.24\%\\
   1.73621$\mu$m & $n=  4 - n=  10$ & & 15 & 20& 1.80\%& 15 & 20& 1.038\% & 15& 15 & 0.93\%\\
   1.81741$\mu$m & $n=  4 - n=  9$ & & 15 & 20& 2.06\%& 15& 20& 1.25\%& 15& 15 & 0.62\%\\  
   1.87510$\mu$m & $n=  3 - n=  4$ & & 15 & 25& 5.46\%& 15& 20& 2.62\% & 15& 15 & 0.51\%\\
   1.94456$\mu$m & $n=  4 - n=  8$ & & 15 & 20& 2.51\%& 15& 20& 1.31\%& 15& 15 & 0.48\%\\  
   2.16553$\mu$m & $n=  4 - n=  7$ & & 15 & 20& 3.30\%& 15& 20& 1.61\%& 15& 15 & 0.45\%\\ 
   2.62515$\mu$m & $n=  4 - n=  6$ & & 15 & 20& 4.89\%& 15& 20& 2.34\%& 15& 15& 0.53\%\\
   3.03837$\mu$m & $n=  5 - n=  10$ & & 15 & 20& 4.14\%& 15 & 20& 2.53\%& 15& 20& 1.13\%\\
   3.29609$\mu$m & $n=  5 - n=  9$ & & 15 & 20& 4.76\%& 15 & 20& 2.56\%& 15& 15& 0.85\%\\
   3.73954$\mu$m & $n=  5 - n=  8$ & & 15 & 25& 5.78\%& 15 & 20& 2.89\%& 15& 15& 0.77\%\\
   4.05115$\mu$m & $n=  4 - n=  5$ & & 20 & 25& 9.14\%& 15& 20& 4.39\%& 15& 15& 0.83\%\\ 
   4.65251$\mu$m & $n=  5 - n=  7$ & & 20 & 25& 7.74\%& 15 & 20& 3.76\%& 15& 15& 0.83\%\\
   5.12726$\mu$m & $n=  6 - n=  10$ & & 20 & 25& 7.05\%& 15 & 20& 3.92\%& 15& 20& 1.36\%\\
   5.90660$\mu$m & $n=  6 - n=  9$ & & 20 & 25& 8.29\%& 15 & 20& 4.26\%& 15& 20& 1.13\%\\
   7.45782$\mu$m & $n=  5 - n=  6$ & & 20 & 25& 12.44\%& 20 & 20& 6.00\%& 15& 20& 1.11\%\\
   7.50045$\mu$m & $n=  6 - n=  8$ & & 20 & 25& 10.50\%& 15&20& 5.17\%& 15& 20& 1.13\%\\
   8.75768$\mu$m & $n=  7 - n=  10$ & & 20 & 25& 10.66\%& 15 &20& 5.66\%& 15& 20& 1.63\%\\
   11.3056$\mu$m & $n=  7 - n=  9$ & & 20 & 25& 13.06\%& 20 & 20& 6.58\%& 15& 20& 1.46\% \\
   12.3685$\mu$m & $n=  6 - n=  7$ & & 20 & 25& 15.35\%& 20 & 20& 7.46\%& 15& 20& 1.35\%\\
   16.2047$\mu$m & $n=  8 - n=  10$ & & 20 & 25& 15.31\%& 20 & 20& 7.99\%& 15& 20& 1.98\%\\
   19.0567$\mu$m & $n=  7 - n=  8$ & & 20 & 25& 17.86\%& 20 & 20& 8.81\%& 15& 20& 1.60\%\\
   27.7958$\mu$m & $n=  8 - n=  9$ & & 25 & 30& 20.01\%& 20 &20& 10.13\%& 15& 20& 1.92\%\\
\end{tabular}
\end{minipage}
 \end{sideways}
\end{table*}

\begin{table*}
    \centering
        \caption{ He I recombination lines 
        tested in our hydrogen and helium 
        mixture model at $T=10^4$K. The columns in the right contain the maximum 
        resolved principal quantum number $n$ 
        for which the line intensity converge 
        for less than 5\% and 1\% with respect 
        to the models with maximum resolved 
        levels $n^\text{res}= n-5$ (convergence ratios start to be calculated at $n=15$). We show these at two 
        relevant electron densities of $\n_\text{H} = 
        10\text{cm}^{-3}$ and $\n_\text{H} = 
        10^2\text{cm}^{-3}$. For each density, the maximum percentage difference induced by changing the maximum resolved level from  $n = 70$ to $n = 10$ is also shown (columns "Diff.").}
    \label{tab:lineshhelowden}
    \begin{tabular}{c|c|c|c|c|c|c|c|c|c}
       transition type & $\lambda$ & transition & comments & \multicolumn{3}{|c|}{Convergence at $\n_\text{H}=10\text{cm}^{-3}$}  & \multicolumn{3}{|c|}{Convergence at $\n_\text{H}=10^2\text{cm}^{-3}$} \\
      &&&& n ($<$5\%) & n ($<$1\%) & Diff. &n ($<$5\%) & n ($<$1\%) & Diff\\
      \hline
      \hline 
    \multirow{19}{*}{singlet-singlet} & 512.099\AA & $1\,^1S -  6\,^1P_1$ & & 15 & 20& 3.09\%& 15 & 20 & 3.04\%\\
    & 515.617\AA & $1\,^1S -  5\,^1P_1$ & & 15 & 20& 3.57\%& 15 & 20 & 3.52\%\\
    & 522.213\AA & $1\,^1S -  4\,^1P_1$ & & 15 & 20& 3.19\%& 15 & 20 & 3.14\%\\
    & 537.030\AA & $1\,^1S -  3\,^1P_1$ & & 15 & 20& 2.38\%& 15 & 20 & 2.32\%\\
    & 584.334\AA & $1\,^1S -  2\,^1P_1$ & & 15 & 15& 1.22\%& 15 & 15 & 1.14\%\\
    & 3447.59\AA & $2\,^1S -  6\,^1P_1$ & & 15 & 20& 3.12\%& 15 & 20 & 3.08\% \\
    & 3964.73\AA & $2\,^1S -  4\,^1P_1$ & & 15 & 20& 3.22\%& 15 & 20 & 3.17\% \\
     & 4143.76\AA & $2\,^1P_1 -   6\,^1D_2$ & & 15 & 15& 0.86\%& 15 & 15 & 0.84\% \\
     & 4168.97\AA & $2\,^1P_1 -   6\,^1S$ & & 15 & 15& 0.60\%& 15 & 15 & 0.60\% \\
     & 4387.93\AA & $2\,^1P_1 -   5\,^1D_2$ && 15 & 15& 0.19\%& 15 & 15 & 0.19\%  \\
     & 4437.55\AA & $2\,^1P_1 -   5\,^1S$ & & 15 & 15& 0.71\%& 15 & 15 & 0.72\% \\
     & 4921.93\AA & $2\,^1P_1 -   4\,^1D_2$ & & 15 & 15& 1.39\%& 15 & 15 & 1.35\% \\
     & 5015.68\AA & $2\,^1S -  3\,^1P_1$ & & 15 & 20& 2.40\%& 15 & 20 & 2.36\% \\
     & 5047.74\AA & $2\,^1P_1 -   4\,^1S$ & & 15 & 15& 0.19\%& 15 & 15 & 0.20\%\\
     & 6678.15\AA & $2\,^1P_1 -   3\,^1D_2$ & & 15 & 25& 7.33\%& 15 & 25 & 7.10\%\\
     & 7281.35\AA & $2\,^1P_1 -   3\,^1S$ & & 15 & 15& 0.60\%& 15 & 15 & 0.57\% \\
     & 9603.44\AA & $3\,^1S -   6\,^1P_1$ & & 15 & 20& 3.12\%& 15 & 20 & 3.08\% \\
     & 12790.5\AA & $3\,^1D_2 -   5\,^1F_3$ & & 15 & 25& 7.99\%& 15 & 25 & 7.77\% \\
     & 20581.3\AA & $2\,^1S -   2\,^1P_1$ & & 15 & 15& 1.22\%& 15 & 15 & 1.13\% \\    
\hline
    \multirow{23}{*}{triplet-triplet} & 2829.08\AA & $2\,^3S -  6\,^3P$ & & 15 & 20& 3.15\%& 15 & 20 & 3.09\%  \\
     & 2945.10\AA & $2\,^3S -  5\,^3P$ & & 15 & 20& 3.99\%& 15 & 20 & 3.93\% \\
     & 3187.74\AA & $2\,^3S -  4\,^3P$ & & 15 & 20& 3.66\%& 15 & 20 & 3.60\% \\
     & 3819.62\AA & $2\,^3P_J -  6\,^3D$ & Blend of $J=0,1,2$  & 15 & 15& 0.14\%& 15 & 15 & 0.12\% \\
     & 3867.49\AA & $2\,^3P_J -  6\,^3S$ & Blend of $J=0,1,2$ & 15 & 25& 5.84\%& 15 & 25 & 5.78\% \\
     & 3888.64\AA & $2\,^3S -  3\,^3P$ & & 15 & 20& 2.94\%& 15 & 20 & 2.88\% \\
     & 4026.21\AA & $2\,^3P_J -  5\,^3D$ & Blend of $J=0,1,2$ & 15 & 15& 0.04\%& 15 & 15 & 0.04\% \\
     & 4120.84\AA & $2\,^3P_J -  5\,^3S$ & Blend of $J=0,1,2$ & 15 & 20& 1.84\%& 15 & 20 & 1.78\% \\
     & 4471.50\AA & $2\,^3P_J -  4\,^3D$ & Blend of $J=0,1,2$ & 15 & 15& 1.39\%& 15 & 15 & 1.36\% \\
     & 4713.17\AA & $2\,^3P_J -  4\,^3S$ & Blend of $J=0,1,2$ & 15 & 15& 0.99\% & 15 & 15 & 0.95\%\\
     & 5875.66\AA & $2\,^3P_J -  3\,^3D$ & Blend of $J=0,1,2$ & 15 & 25& 7.05\%& 15 & 25 & 6.83\% \\
     & 7065.25\AA & $2\,^3P_J -  3\,^3S$ & Blend of $J=0,1,2$ & 15 & 15& 0.47\%& 15 & 15 & 0.46\%  \\
    & 8361.73\AA & $3\,^3S -  6\,^3P$ & & 15 & 20& 3.15\%& 15 & 20 & 3.09\% \\
     & 9463.58\AA & $3\,^3S -   5\,^3P$ & & 15 & 20& 3.98\%& 15 & 20 & 3.93\% \\
     & 10830.2\AA & $2\,^3S -   2\,^3P_J$ & Blend of $J=0,1,2$ & 15 & 15& 0.35\%& 15 & 15 & 0.27\% \\
     & 10913.0\AA & $3\,^3D -   6\,^3F$ & & 15 & 25& 5.05\%& 15 & 25 & 4.94\% \\
     & 12527.5\AA & $3\,^3S -   4\,^3P$ & & 15 & 20& 3.66\%& 15 & 20 & 3.60\% \\
     & 12784.9\AA & $3\,^3D -   5\,^3F$ & & 15 & 25& 7.98\%& 15 & 25 & 7.78\% \\
     & 12846.0\AA & $3\,^3P -   5\,^3S$ & & 15 & 20& 1.84\%& 15 & 20 & 1.78\% \\
     & 12984.9\AA & $3\,^3D -   5\,^3P$ & & 15 & 20& 3.99\%& 15 & 20 & 3.93\% \\
     & 18685.4\AA & $3\,^3D -   4\,^3F$ && 20 & 35& 16.12\%& 20 & 35 & 15.66\%  \\
     & 19543.1\AA & $3\,^3D -   4\,^3P$ & & 15 & 20& 3.66\%& 15 & 20 & 3.60\% \\
    & 21120.2\AA & $3\,^3P -   4\,^3S$ && 15 & 15& 1.00\%& 15 & 15 & 0.94\%  \\
    \hline
        \multirow{3}{*}{singlet-triplet}& 591.409\AA & $1\,^1S -  2\,^3P_0$ & & 15 & 15& 0.35\%& 15 & 15 & 0.30\% \\
    & 625.563\AA & $1\,^1S -  2\,^3S$ & & 15 & 15& 0.16\%& 15 & 15 & 0.14\%\\
     & 8863.66\AA & $2\,^3S -  2\,^1P_1$ & & 15 & 15& 1.22\%& 15 & 15 & 1.13\% \\
\end{tabular}
\end{table*}

\begin{table*}
    \centering
        \caption{ He I recombination lines 
        tested in our hydrogen and helium 
        mixture model at $T=10^4$K. The columns in the right contain the maximum 
        resolved principal quantum number $n$ 
        for which the line intensity converge 
        for less than 5\% and 1\% with respect 
        to the models with maximum resolved 
        levels $n^\text{res}= n-5$ (convergence ratios start to be calculated at $n=15$). We show these at electron densities of $\n_\text{H} = 
        10^3\text{cm}^{-3}$ and $\n_\text{H} = 
        10^5\text{cm}^{-3}$. For each density, the maximum percentage difference induced by changing the maximum resolved level from  $n = 70$ to $n = 10$ is also shown (columns "Diff.").}
    \label{tab:lineshheintden}
    \begin{tabular}{c|c|c|c|c|c|c|c|c|c}
       transition type & $\lambda$ & transition & comments & \multicolumn{3}{|c|}{Convergence at $\n_\text{H}=10^3\text{cm}^{-3}$}  & \multicolumn{3}{|c|}{Convergence at $\n_\text{H}=10^5\text{cm}^{-3}$} \\
      &&&& n ($<$5\%) & n ($<$1\%) & Diff. &n ($<$5\%) & n ($<$1\%) & Diff\\
      \hline
      \hline 
    \multirow{19}{*}{singlet-singlet} & 512.099\AA & $1\,^1S -  6\,^1P_1$ & & 15 & 20& 2.92\%& 15 & 20 & 2.19\%\\
    & 515.617\AA & $1\,^1S -  5\,^1P_1$ & & 15 & 20& 3.36\%& 15 & 20 & 2.59\%\\
    & 522.213\AA & $1\,^1S -  4\,^1P_1$ & & 15 & 20& 2.99\%& 15 & 20 & 2.27\%\\
    & 537.030\AA & $1\,^1S -  3\,^1P_1$ & & 15 & 20& 2.19\%& 15 & 15 & 1.61\%\\
    & 584.334\AA & $1\,^1S -  2\,^1P_1$ & & 15 & 15& 0.86\%& 15 & 15 & 0.36\%\\
    & 3447.59\AA & $2\,^1S -  6\,^1P_1$ & & 15 & 20& 2.96\%& 15 & 20 & 2.22\% \\
    & 3964.73\AA & $2\,^1S -  4\,^1P_1$ & & 15 & 20& 3.03\%& 15 & 20 & 2.30\% \\
     & 4143.76\AA & $2\,^1P_1 -   6\,^1D_2$ & & 15 & 15& 0.80\%& 15 & 15 & 0.48\% \\
     & 4168.97\AA & $2\,^1P_1 -   6\,^1S$ & & 15 & 15& 0.63\%& 15 & 15 & 0.41\% \\
     & 4387.93\AA & $2\,^1P_1 -   5\,^1D_2$ && 15 & 15& 0.17\%& 15 & 15 & 0.01\%  \\
     & 4437.55\AA & $2\,^1P_1 -   5\,^1S$ & & 15 & 15& 0.73\%& 15 & 15 & 0.56\% \\
     & 4921.93\AA & $2\,^1P_1 -   4\,^1D_2$ & & 15 & 15& 1.26\%& 15 & 15 & 1.01\% \\
     & 5015.68\AA & $2\,^1S -  3\,^1P_1$ & & 15 & 20& 2.23\%& 15 & 20 & 1.64\% \\
     & 5047.74\AA & $2\,^1P_1 -   4\,^1S$ & & 15 & 15& 0.23\%& 15 & 15 & 0.19\%\\
     & 6678.15\AA & $2\,^1P_1 -   3\,^1D_2$ & & 15 & 25& 6.51\%& 15 & 25 & 4.51\%\\
     & 7281.35\AA & $2\,^1P_1 -   3\,^1S$ & & 15 & 15& 0.48\%& 15 & 15 & 0.28\% \\
     & 9603.44\AA & $3\,^1S -   6\,^1P_1$ & & 15 & 20& 2.96\%& 15 & 20 & 2.22\% \\
     & 12790.5\AA & $3\,^1D_2 -   5\,^1F_3$ & & 15 & 25& 7.30\%& 15 & 25 & 5.37\% \\
     & 20581.3\AA & $2\,^1S -   2\,^1P_1$ & & 15 & 15& 0.85\%& 15 & 15 & 0.35\% \\    
\hline
    \multirow{23}{*}{triplet-triplet} & 2829.08\AA & $2\,^3S -  6\,^3P$ & & 15 & 20& 2.94\%& 15 & 20 & 2.10\%  \\
     & 2945.10\AA & $2\,^3S -  5\,^3P$ & & 15 & 20& 3.75\%& 15 & 20 & 2.88\% \\
     & 3187.74\AA & $2\,^3S -  4\,^3P$ & & 15 & 20& 3.39\%& 15 & 20 & 2.50\% \\
     & 3819.62\AA & $2\,^3P_J -  6\,^3D$ & Blend of $J=0,1,2$  & 15 & 15& 0.06\%& 15 & 15 & 0.25\% \\
     & 3867.49\AA & $2\,^3P_J -  6\,^3S$ & Blend of $J=0,1,2$ & 15 & 25& 5.66\%& 15 & 20 & 5.14\% \\
     & 3888.64\AA & $2\,^3S -  3\,^3P$ & & 15 & 20& 2.65\%& 15 & 20 & 1.78\% \\
     & 4026.21\AA & $2\,^3P_J -  5\,^3D$ & Blend of $J=0,1,2$ & 15 & 15& 0.08\%& 15 & 15 & 0.26\% \\
     & 4120.84\AA & $2\,^3P_J -  5\,^3S$ & Blend of $J=0,1,2$ & 15 & 20& 1.59\%& 15 & 15 & 1.02\% \\
     & 4471.50\AA & $2\,^3P_J -  4\,^3D$ & Blend of $J=0,1,2$ & 15 & 15& 1.31\%& 15 & 15 & 1.14\% \\
     & 4713.17\AA & $2\,^3P_J -  4\,^3S$ & Blend of $J=0,1,2$ & 15 & 15& 0.83\% & 15 & 15 & 0.49\%\\
     & 5875.66\AA & $2\,^3P_J -  3\,^3D$ & Blend of $J=0,1,2$ & 15 & 25& 6.26\%& 15 & 25 & 4.30\% \\
     & 7065.25\AA & $2\,^3P_J -  3\,^3S$ & Blend of $J=0,1,2$ & 15 & 15& 0.38\%& 15 & 15 & 0.21\%  \\
    & 8361.73\AA & $3\,^3S -  6\,^3P$ & & 15 & 20& 2.10\%& 15 & 20 & 2.94\% \\
     & 9463.58\AA & $3\,^3S -   5\,^3P$ & & 15 & 20& 3.75\%& 15 & 20 & 2.88\% \\
     & 10830.2\AA & $2\,^3S -   2\,^3P_J$ & Blend of $J=0,1,2$ & 15 & 15& 0.09\%& 15 & 15 & 0.02\% \\
     & 10913.0\AA & $3\,^3D -   6\,^3F$ & & 15 & 25& 4.76\%& 15 & 20 & 3.88\% \\
     & 12527.5\AA & $3\,^3S -   4\,^3P$ & & 15 & 20& 3.39\%& 15 & 20 & 2.50\% \\
     & 12784.9\AA & $3\,^3D -   5\,^3F$ & & 15 & 25& 7.38\%& 15 & 25 & 5.63\% \\
     & 12846.0\AA & $3\,^3P -   5\,^3S$ & & 15 & 20& 1.59\%& 15 & 15 & 1.02\% \\
     & 12984.9\AA & $3\,^3D -   5\,^3P$ & & 15 & 20& 3.75\%& 15 & 20 & 2.88\% \\
     & 18685.4\AA & $3\,^3D -   4\,^3F$ && 20 & 35& 14.70\%& 20 & 25 & 10.72\%  \\
     & 19543.1\AA & $3\,^3D -   4\,^3P$ & & 15 & 20& 3.40\%& 15 & 20 & 2.50\% \\
    & 21120.2\AA & $3\,^3P -   4\,^3S$ && 15 & 15& 0.83\%& 15 & 15 & 0.49\%  \\
    \hline
        \multirow{3}{*}{singlet-triplet}& 591.409\AA & $1\,^1S -  2\,^3P_0$ & & 15 & 15& 0.20\%& 15 & 15 & 0.07\% \\
    & 625.563\AA & $1\,^1S -  2\,^3S$ & & 15 & 15& 0.16\%& 15 & 15 & 0.11\%\\
     & 8863.66\AA & $2\,^3S -  2\,^1P_1$ & & 15 & 15& 0.85\%& 15 & 15 & 0.35\% \\
\end{tabular}
\end{table*}

\begin{table*}
    \centering
        \caption{ He I recombination lines 
        tested in our hydrogen and helium 
        mixture model at $T=10^4$K. The columns in the right contain the maximum 
        resolved principal quantum number $n$ 
        for which the line intensity converge 
        for less than 5\% and 1\% with respect 
        to the models with maximum resolved 
        levels $n^\text{res}= n-5$ (convergence ratios start to be calculated at $n=15$). We show these at two 
        relevant electron densities of $\n_\text{H} = 
        10^6\text{cm}^{-3}$ and $\n_\text{H} = 
        10^7\text{cm}^{-3}$. For each density, the maximum percentage difference induced by changing the maximum resolved level from  $n = 70$ to $n = 10$ is also shown (columns "Diff.").}
    \label{tab:lineshhehighden}
    \begin{tabular}{c|c|c|c|c|c|c|c|c|c}
       transition type & $\lambda$ & transition & comments & \multicolumn{3}{|c|}{Convergence at $\n_\text{H}=10^6\text{cm}^{-3}$}  & \multicolumn{3}{|c|}{Convergence at $\n_\text{H}=10^7\text{cm}^{-3}$} \\
      &&&& n ($<$5\%) & n ($<$1\%) & Diff. &n ($<$5\%) & n ($<$1\%) & Diff\\
      \hline
      \hline 
    \multirow{19}{*}{singlet-singlet} & 512.099\AA & $1\,^1S -  6\,^1P_1$ & & 15 & 20& 1.33\%& 15 & 15 & 0.23\%\\
    & 515.617\AA  & $1\,^1S -  5\,^1P_1$ & & 15 & 20& 1.80\%& 15 & 15 & 0.83\%\\
    & 522.213\AA  & $1\,^1S -  4\,^1P_1$ & & 15 & 20& 1.57\%& 15 & 15 & 0.72\%\\
    & 537.030\AA  & $1\,^1S -  3\,^1P_1$ & & 15 & 15& 1.09\%& 15 & 15 & 0.50\%\\
    & 584.334\AA  & $1\,^1S -  2\,^1P_1$ & & 15 & 15& 0.23\%& 15 & 15 & 0.03\%\\
    & 3447.59\AA  & $2\,^1S -  6\,^1P_1$ & & 15 & 20& 1.35\%& 15 & 15 & 0.24\% \\
    & 3964.73\AA  & $2\,^1S -  4\,^1P_1$ & & 15 & 20& 1.59\%& 15 & 15 & 0.73\% \\
     & 4143.76\AA  & $2\,^1P_1 -   6\,^1D_2$ & & 15 & 15& 0.22\%& 15 & 15 & 0.09\% \\
     & 4168.97\AA  & $2\,^1P_1 -   6\,^1S$ & & 15 & 15& 0.03\%& 15 & 15 & 0.59\% \\
     & 4387.93\AA  & $2\,^1P_1 -   5\,^1D_2$ && 15 & 15& 0.12\%& 15 & 15 & 0.09\%  \\
     & 4437.55\AA  & $2\,^1P_1 -   5\,^1S$ & & 15 & 15& 0.28\%& 15 & 15 & 0.05\% \\
     & 4921.93\AA  & $2\,^1P_1 -   4\,^1D_2$ & & 15 & 15& 0.81\%& 15 & 15 & 0.36\% \\
     & 5015.68\AA  & $2\,^1S -  3\,^1P_1$ & & 15 & 15& 1.11\%& 15 & 15 & 0.51\% \\
     & 5047.74\AA  & $2\,^1P_1 -   4\,^1S$ & & 15 & 15& 0.05\%& 15 & 15 & 0.12\%\\
     & 6678.15\AA  & $2\,^1P_1 -   3\,^1D_2$ & & 15 & 20& 3.06\%& 15 & 20 & 1.14\%\\
     & 7281.35\AA  & $2\,^1P_1 -   3\,^1S$ & & 15 & 15& 0.24\%& 15 & 15 & 0.18\% \\
     & 9603.44\AA  & $3\,^1S -   6\,^1P_1$ & & 15 & 20& 1.35\%& 15 & 15 & 0.24\% \\
     & 12790.5\AA  & $3\,^1D_2 -   5\,^1F_3$ & & 15 & 20& 3.71\%& 15 & 20 & 1.54\% \\
     & 20581.3\AA  & $2\,^1S -   2\,^1P_1$ & & 15 & 15& 0.22\%& 15 & 15 & 0.03\% \\    
\hline
    \multirow{23}{*}{triplet-triplet} & 2829.08\AA  & $2\,^3S -  6\,^3P$ & & 15 & 20& 1.13\%& 15 & 15 & 0.25\%  \\
     & 2945.10\AA  & $2\,^3S -  5\,^3P$ & & 15 & 20& 2.03\%& 15 & 15 & 0.85\% \\
     & 3187.74\AA  & $2\,^3S -  4\,^3P$ & & 15 & 20& 1.75\%& 15 & 15 & 0.73\% \\
     & 3819.62\AA  & $2\,^3P_J -  6\,^3D$ & Blend of $J=0,1,2$  & 15 & 15& 0.45\%& 15 & 15 & 0.52\% \\
     & 3867.49\AA  & $2\,^3P_J -  6\,^3S$ & Blend of $J=0,1,2$ & 15 & 20& 4.64\%& 15 & 20 & 4.14\% \\
     & 3888.64\AA  & $2\,^3S -  3\,^3P$ & & 15 & 20& 1.23\%& 15 & 15 & 0.50\% \\
     & 4026.21\AA  & $2\,^3P_J -  5\,^3D$ & Blend of $J=0,1,2$ & 15 & 15& 0.34\%& 15 & 15 & 0.29\% \\
     & 4120.84\AA  & $2\,^3P_J -  5\,^3S$ & Blend of $J=0,1,2$ & 15 & 15& 0.67\%& 15 & 15 & 0.25\% \\
     & 4471.50\AA  & $2\,^3P_J -  4\,^3D$ & Blend of $J=0,1,2$ & 15 & 15& 0.93\%& 15 & 15 & 0.53\% \\
     & 4713.17\AA  & $2\,^3P_J -  4\,^3S$ & Blend of $J=0,1,2$ & 15 & 15& 0.32\% & 15 & 15 & 0.13\%\\
     & 5875.66\AA  & $2\,^3P_J -  3\,^3D$ & Blend of $J=0,1,2$ & 15 & 20& 2.97\%& 15 & 20 & 1.25\% \\
     & 7065.25\AA  & $2\,^3P_J -  3\,^3S$ & Blend of $J=0,1,2$ & 15 & 15& 0.15\%& 15 & 15 & 0.05\%  \\
    & 8361.73\AA  & $3\,^3S -  6\,^3P$ & & 15 & 20& 1.12\%& 15 & 15 & 0.26\% \\
     & 9463.58\AA  & $3\,^3S -   5\,^3P$ & & 15 & 20& 2.03\%& 15 & 15 & 0.85\% \\
     & 10830.2\AA  & $2\,^3S -   2\,^3P_J$ & Blend of $J=0,1,2$ & 15 & 15& 0.01\%& 15 & 15 & 0.01\% \\
     & 10913.0\AA  & $3\,^3D -   6\,^3F$ & & 15 & 25& 2.88\%& 15 & 20 & 1.46\% \\
     & 12527.5\AA  & $3\,^3S -   4\,^3P$ & & 15 & 20& 1.75\%& 15 & 15 & 0.73\% \\
     & 12784.9\AA  & $3\,^3D -   5\,^3F$ & & 15 & 20& 4.01\%& 15 & 20 & 1.86\% \\
     & 12846.0\AA  & $3\,^3P -   5\,^3S$ & & 15 & 15& 0.67\%& 15 & 15 & 0.25\% \\
     & 12984.9\AA  & $3\,^3D -   5\,^3P$ & & 15 & 20& 2.02\%& 15 & 15 & 0.85\% \\
     & 18685.4\AA  & $3\,^3D -   4\,^3F$ && 20 & 25& 7.17\%& 20 & 20 & 2.84\%  \\
     & 19543.1\AA  & $3\,^3D -   4\,^3P$ & & 15 & 20& 1.75\%& 15 & 15 & 0.74\% \\
    & 21120.2\AA  & $3\,^3P -   4\,^3S$ && 15 & 15& 0.32\%& 15 & 15 & 0.13\%  \\
    \hline
        \multirow{3}{*}{singlet-triplet}& 591.409\AA  & $1\,^1S -  2\,^3P_0$ & & 15 & 15& 0.05\%& 15 & 15 & 0.03\% \\
    & 625.563\AA  & $1\,^1S -  2\,^3S$ & & 15 & 15& 0.07\%& 15 & 15 & 0.02\%\\
     & 8863.66\AA  & $2\,^3S -  2\,^1P_1$ & & 15 & 15& 0.22\%& 15 & 15 & 0.03\% \\
\end{tabular}
\end{table*}


\bsp	
\label{lastpage}
\end{document}